\documentclass[pra,aps,amsmath,amssymb,amsfonts,twocolumn,nofootinbib,longbibliography,superscriptaddress]{revtex4-1}
\usepackage{amssymb}
\usepackage{bm,mathrsfs}
\usepackage{graphicx}
\usepackage{epsfig}
\usepackage{amsmath,bbm}
\usepackage{amsfonts,amssymb}
\usepackage{times}
\usepackage{verbatim}
\usepackage[sort&compress]{natbib}
\usepackage{amsmath}
\usepackage{bm}
\usepackage[colorlinks,breaklinks,linkcolor=blue,anchorcolor=blue,citecolor=blue,urlcolor=blue]{hyperref}

\begin{document}

\title{Coherent ground-state transport of neutral atoms}

\author{X. X. Li}
\affiliation{Center for Quantum Sciences and School of Physics, Northeast Normal University, Changchun 130024,  China}
\author{J. B. You}
\affiliation{Institute of High Performance Computing, A*STAR (Agency for Science, Technology and Research), 1 Fusionopolis Way, Connexis, Singapore 138632}
\author{X. Q. Shao}
\email{shaoxq644@nenu.edu.cn}
\affiliation{Center for Quantum Sciences and School of Physics, Northeast Normal University, Changchun 130024, China}
\affiliation{Center for Advanced Optoelectronic Functional Materials Research, and Key Laboratory for UV Light-Emitting Materials and Technology
of Ministry of Education, Northeast Normal University, Changchun 130024, China}
\author{Weibin Li}
\affiliation{School of Physics and Astronomy, The University of Nottingham, Nottingham NG7 2RD, United Kingdom}

\begin{abstract}
Quantum state transport is an important way to study the energy or information flow. By combining the unconventional Rydberg pumping mechanism and the diagonal form of van der Waals interactions, we construct a theoretical model via second-order perturbation theory to realize a long-range coherent transport inside the ground-state manifold of neutral atoms system. With the adjustment of the Rabi frequencies and the interatomic
distance, this model can be used to simulate various single-body physics phenomena such as Heisenberg $XX$ spin chain restricted in the single-excitation manifold, coherently perfect quantum state transfer, parameter adjustable Su-Schrieffer-Heeger model, and chiral motion of atomic excitation in the triangle by varying the geometrical
arrangement of the three atoms, which effectively avoid the influence of atomic spontaneous emission at the same time. Moreover, the influence of atomic position fluctuation on the fidelity of quantum state transmission is discussed in detail, and the corresponding numerical results show that our work provides a robust and easy-implemented scheme for quantum state transport with neutral atoms.
\end{abstract}

\maketitle

\section{Introduction}
Quantum state transport plays an important role in understanding the energy or information flow at the microscopic particle level. Because of its simplicity, spin-chain system with nearest-neighbor hopping has been extensively used to realize quantum state transmission \cite{PhysRevLett.91.207901,PhysRevA.72.034303,PhysRevA.91.042321,PhysRevB.94.045314,PhysRevA.75.050303,PhysRevA.69.034304,PhysRevLett.99.250506,PhysRevA.95.012317,PhysRevLett.93.230502,doi:10.1080/00107510701342313}.  In order to achieve high-fidelity transfer of quantum information, various transport protocols have been put forward, such as modulation of the couplings between neighboring spins \cite{PhysRevLett.92.187902,PhysRevA.72.030301,PhysRevLett.97.180502,PhysRevA.71.032310,2019Perfect,PhysRevA.85.012323}, exploitation of the chiral topological edge states \cite{Dlaska_2017,ncomms2531}, and construction of a stimulated
Raman adiabatic passage \cite{PhysRevB.70.235317,PhysRevA.90.012319,Eckert_2007}, especially combined with the topologically protected edge states \cite{PhysRevA.98.012331,PhysRevResearch.2.033475}. Among many physical systems, Rydberg atom has been regarded as a good candidate to simulate spin-chain models on account of its remarkable properties \cite{PhysRevX.8.021070,nature18274,Browaeys2020,nature24622,RevModPhys.82.2313,Whitlock_2017,AQS_2021}. In particular, the long-range interactions are capable of causing diverse consequences such as Rydberg blockade \cite{PhysRevLett.87.037901,PhysRevLett.99.073002,nphys1178} and antiblockade \cite{PhysRevLett.98.023002,PhysRevLett.104.013001,PhysRevLett.118.063606,PhysRevA.99.060101,PhysRevLett.125.033602} over long-range molecules \cite{PhysRevLett.85.2458,PhysRevLett.102.173001}.

Recently quantum state transfer schemes based on Rydberg atoms have made rapid progresses both theoretically and experimentally  \cite{PhysRevLett.114.113002,PhysRevLett.114.123005,PhysRevLett.120.063601,2007Motion,PhysRevLett.115.093002,PhysRevA.97.043415,
PhysRevLett.123.063001,science.1244843,NewJ.Phys.13.073044,PhysRevLett.114.173002,PhysRevLett.114.243002,nature13461}. According to different coding modes of qubit, these schemes can be divided into three categories. One is the spin-exchange between Rydberg states \cite{PhysRevLett.114.113002,PhysRevLett.114.123005,PhysRevLett.120.063601,2007Motion}. For instance, Barredo {\it et al.} \cite{PhysRevLett.114.113002} studied this hopping in a spin chain constructed by individually addressable Rydberg atoms utilizing the long-range resonant dipole-dipole coupling. The second one is the quantum state transfer between ground state and Rydberg state which remains as a second-order process in terms of laser-spin coupling \cite{PhysRevLett.115.093002,PhysRevA.97.043415,
PhysRevLett.123.063001,science.1244843}. To reach this target, Yang {\it et al.} \cite{PhysRevLett.123.063001} constructed an exchange interaction between ground state and Rydberg state, mediated by synthetic spin exchange arising from diagonal van der Waals (vdW) interaction. The last one is the excitation transport taking place in the ground-state manifold through a fourth-order process   \cite{NewJ.Phys.13.073044,PhysRevLett.114.173002,PhysRevLett.114.243002}, where the effective spin-spin interactions between ground state atoms are obtained by dressing Rydberg states with dipole-dipole interaction, vdW interactions, and F\"{o}rster-resonance interaction.

In this work, we make use of the diagonal vdW interactions and the unconventional Rydberg pumping \cite{PhysRevA.98.062338,PhysRevA.102.053118} to realize coherent excitation transport inside ground-state manifold of a series of three-level Rydberg atoms. The simple energy level structure can help reduce the complexity of the experiment operation. Because the evolution dynamics of the whole system is a second-order process, we can easily modulate the effective coupling strength between adjacent sites. Consequently, the current system can be used to simulate various single-body physics phenomena such as Heisenberg $XX$ spin chain restricted in the single-excitation manifold \cite{ncomms13070,Smith2019,Jepsen2020}, coherently perfect quantum state transfer, and parameter adjustable Su-Schrieffer-Heeger (SSH) model \cite{PhysRevLett.42.1698,PhysRevLett.62.2747,JK2016A} by rearrangement of the atoms. The advantage of our system is that no-fine tuning of atomic position is required because the deviation from the unconventional Rydberg pumping condition will only alter the evolution period of the quantum state without destroying the realization of the scheme.

In addition, focusing on three Rydberg atoms arranged in an equilateral triangle, a chiral motion of atomic excitation can be achieved by periodically switching on and off the weak driving fields \cite{nphys3930}, where the excitation hops from site to site in a preferred direction induced by a synthetic gauge field, breaking the time-reversal symmetry. Compared with the recent experimental observation of chiral motion in spin-orbit coupled Rydberg system \cite{PhysRevX.10.021031}, our scheme does not require precise control of electric or magnetic fields.

The remainder of this paper is organized as follows. In Sec.~\ref{II} we introduce the details of the system and derive its effective dynamics via effective operator method. In Sec.~\ref{III} we provide a protocol to realize a one-dimensional SSH model by regulating the detuning of the strong driving field. In Sec.~\ref{IV}, the chiral motion of atomic excitation in the triangle is accomplished via floquet driving. In Sec.~\ref{dis} we further analyze the feasibility of our scheme by considering a realistic experimental setup from multiple perspectives and finally give a conclusion in Sec.~\ref{VI}.
\section{1D chain of atoms.}\label{II}
\begin{figure}
\centering
\includegraphics[scale=0.23]{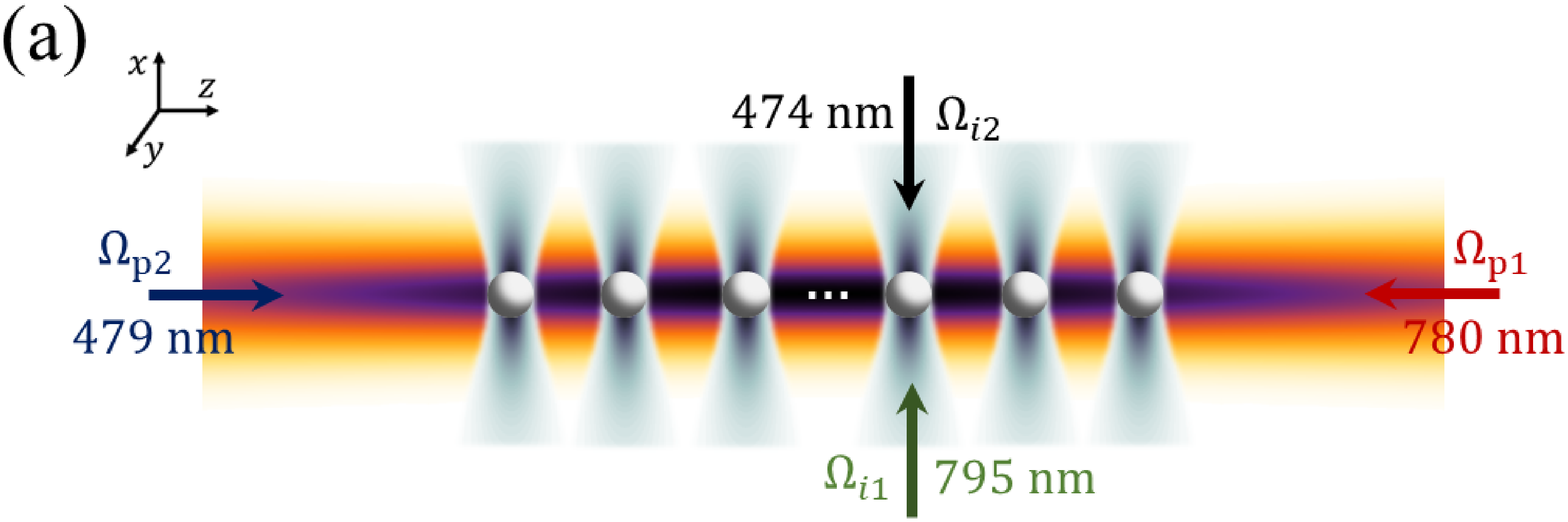}
\hspace{1in}
\includegraphics[scale=0.19]{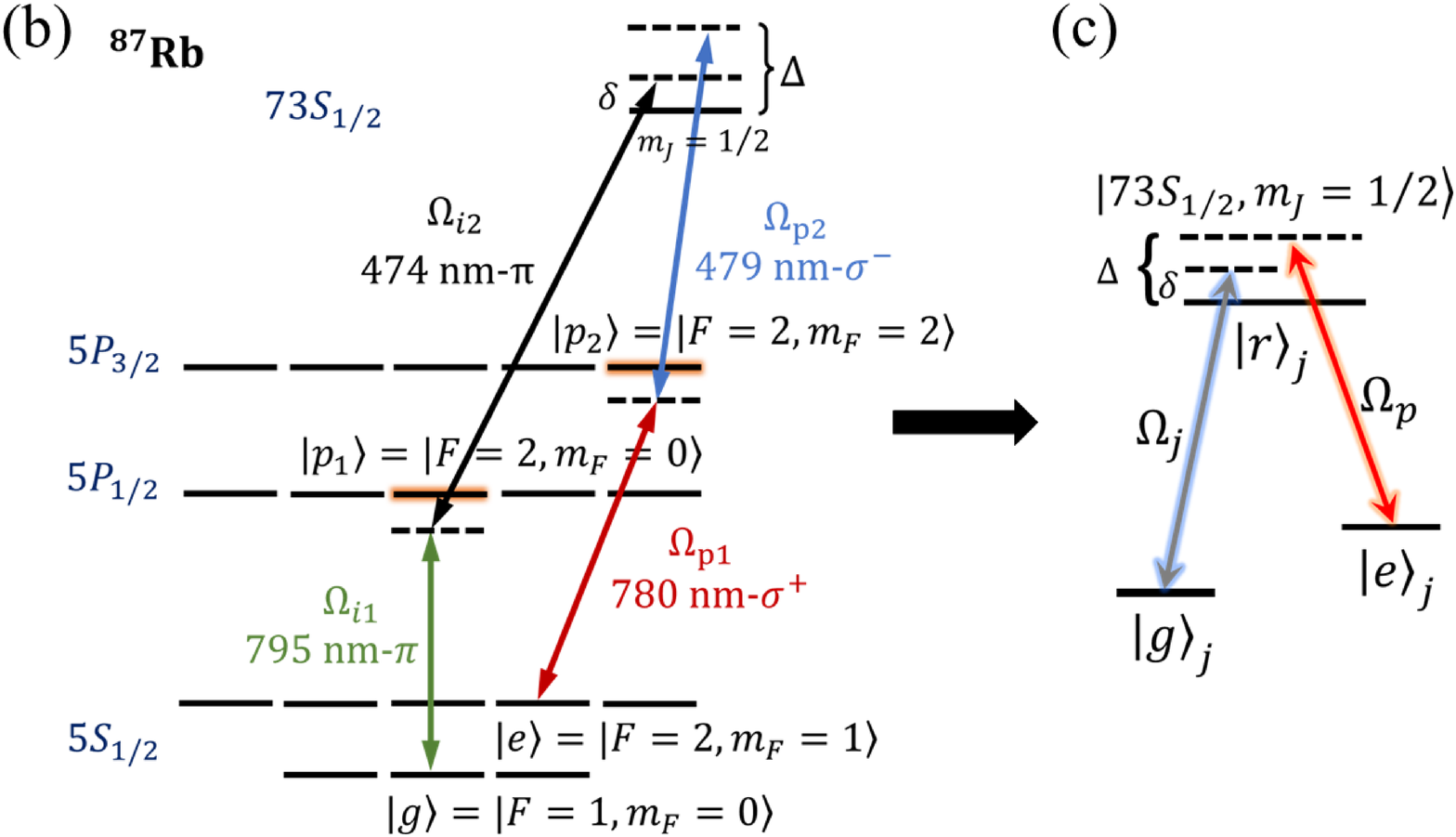}
\caption{\label{ple}Experimental setup. (a) $N$ Rydberg atoms arranged as a linear chain at equal intervals with same energy-level configuration. Atoms are driven by two global laser fields propagating along $z$-axis from two sides. Meanwhile, local laser fields $\Omega_{i1}$ and $\Omega_{i2}$ propagating along $x$-axis are focused onto individual sites. (b) Level structure for the proposed atomic system. We consider $^{87}$Rb with $|g\rangle=|5S_{1/2},F=1,m_{F}=0\rangle$, $|e\rangle=|5S_{1/2},F=2,m_{F}=1\rangle$ and  $|r\rangle=|73S_{1/2},m_{J}=1/2\rangle$. (c) The equivalent system after adiabatically eliminating the intermediate states.}
\end{figure}
The considered system is illustrated in Fig.~\ref{ple}(a), it incorporates $N$ Rydberg atoms ($^{87}$Rb) with the same configuration arranged as a linear chain at equal intervals. The position of the $j$th atom is labeled as $r_{j}$ and the distance between the $j$th and $k$th atoms is $r_{j,k}=|r_{j}-r_{k}|$.
The configuration of each atom is shown in Fig.~\ref{ple}(b).
The ground states here are chosen as $|g\rangle=|5S_{1/2},F=1,m_{F}=0\rangle$, $|e\rangle=|5S_{1/2},F=2,m_{F}=1\rangle$ and the Rydberg state is selected as $|r\rangle=|73S_{1/2},m_{J}=1/2\rangle$.
The transition between ground states and the Rydberg state $|r\rangle$ are driven by two-photon processes,
where state $|g\rangle$ is individually addressed to the intermediate state $|p_{1}\rangle=|5P_{1/2},F=2,m_{F}=0\rangle$ with a $\pi$-polarized laser field $\Omega_{i1}$ at $795~$nm, and then coupled to $|r\rangle$ with another local  $\pi$-polarized dressing laser $\Omega_{i2}$ at $474~$nm,
state $|e\rangle$ is driven by two global laser fields propagating from two sides, where $\Omega_{p1}$ at $780~$nm with $\sigma^{+}$-polarized drives $|e\rangle$ to the intermediate state $|p_{2}\rangle=|5P_{3/2},F=2,m_{F}=2\rangle$ and $\Omega_{p2}$ at $479~$nm with $\sigma^{-}$-polarized coupled $|p_{2}\rangle$ to the Rydberg state $|r\rangle$ \cite{NaturePhysics41567-020-0917-6,nphys2413}.
After adiabatically eliminating the intermediate states $|p_{1(2)}\rangle$, the configuration of each atom can be simplified as a three-level structure shown in Fig.~\ref{ple}(c).
The ground state $|g\rangle$ is dispersively coupled to $|r\rangle$ by a laser field of effective Rabi frequency $\Omega_{j}$ at site $j$, detuning $\delta$, while the transition between $|e\rangle$ and $|r\rangle$ is driven by a global laser field of effective Rabi frequency $\Omega_{p}$, detuned by $\Delta$.
In a rotating frame with respect to $U=\exp[-i\sum_{j=1}^N(\delta|g_j\rangle\langle g_j|+\Delta|e_j\rangle\langle e_j|)t]$, the Hamiltonian of the system reads ($\hbar$ = 1)
\begin{eqnarray}\label{H1}
H_{I}^{(N)}&=&\sum_{j=1}^{N}\Omega_j|r_j\rangle\langle g_j|+\Omega_{p}|r_j\rangle\langle e_j|+{\rm H.c.}+\delta|g_{j}\rangle\langle g_j|\nonumber\\&&+\Delta|e_{j}\rangle\langle e_j|+\sum_{j<k}{\cal U}_{j k}|r_jr_k\rangle\langle r_jr_k|.
\end{eqnarray}
It should be noticed that, to be more intuitive, the phases caused by the wave-vectors have been ignored here, but will be discussed in Sec.~\ref{5a}.
The vdW interaction between atoms in the Rydberg state spaced $r_{j,k}$ takes the form of ${\cal U}_{j k}=C_{6}/r_{j,k}^{6}$.
The second-order non-degenerate  perturbation theory gives that the  dispersion coefficient $C_{6}$ of the vdW interaction is about $1.416~\textmd{THz}\cdot\mu \textmd{m}^{6}$ for state $|73S_{1/2}\rangle$ \cite{SIBALIC2017319}.
So the vdW interaction continuously varies from $2\pi\times1943~$MHz to $2\pi\times1.416~$MHz with $r_{j,j+1}$ adjusted from $3~\mu$m to $10~\mu$m.
Unless otherwise specified, we assume that distance between the nearest neighbor atoms $r_{j,j+1}=4.1~\mu$m for the following numerical simulation, which corresponds to ${\cal U}_{j,j+1}\simeq2\pi\times300~$MHz.

In the limit of large detuning $\Delta\gg\Omega_{p}$ and the unconventional Rydberg pumping condition ${\cal U}_{j,j+1}=\Delta$, the high-frequency oscillating terms proportional to $\Delta$ can be neglected and the computational space is reduced for an initial state $|egg...g\rangle$. Meanwhile, the limiting condition $\{\Omega_{p},\delta\}\gg\Omega_{j}$ allows us to further adiabatically eliminate the Rydberg states via the effective operator method to obtain the effective dynamics of the system \cite{PhysRevA.85.032111,PhysRevLett.106.090502}.

\subsection{Two-atom case}

In order to explain the physical mechanism of the proposed model more clearly, we take the case of $N=2$ as an example and assume $\Omega_{1}=\Omega_{2}=\Omega$, then the Hamiltonian can be simplified as
\begin{eqnarray}
H_{I}^{(2)}&=&\Omega_{p}(|er\rangle\langle rr|+|re\rangle\langle rr|)+\Omega(|re\rangle\langle ge|+|er\rangle\langle eg|)\nonumber\\&&+{\rm H.c.}+\delta(|eg\rangle\langle eg|+|ge\rangle\langle ge|)
\end{eqnarray}
if the system is initialized in the state $|eg\rangle$, where the high-frequency oscillating terms have be neglected under the condition ${\cal U}_{12}=\Delta$ and {$\Delta\gg\Omega_{p}$}.
After shifting the levels of states in this subspace to make the energy of ground states $|ge\rangle$ and $|eg\rangle$ become zero, the Hamiltonian can be further rewritten as $H=H_{0}+V_{+}+V_{-}$, where
\begin{eqnarray}\label{H21}
H_{0}&=&\Omega_{p}(|er\rangle\langle rr|+|re\rangle\langle rr|)+{\rm H.c.}-\delta(|er\rangle\langle er|\nonumber\\&&+|re\rangle\langle re|+|rr\rangle\langle rr|),
\end{eqnarray}
and
\begin{equation}\label{HV}
V_{+}=V^{\dag}_{-}=\Omega(|re\rangle\langle ge|+|er\rangle\langle eg|).
\end{equation}
Here $V_{+}(V_{-})$ are assumed to be perturbative terms under the condition $\Omega\ll\{\Omega_{p},\delta\}$.
Generally speaking, the calculation of dissipation is very complicated if the fine structure of the system is adopted, so we first make a simple assumption that the Rydberg state decays directly into $|g\rangle$ and $|e\rangle$ with the same branching ratio of the spontaneous emission rate, i.e. $\emph{L}_{1}=\sqrt{\gamma/2}|g_1\rangle\langle r_1|$, $\emph{L}_{2}=\sqrt{\gamma/2}|g_2\rangle\langle r_2|$, $\emph{L}_{3}=\sqrt{\gamma/2}|e_1\rangle\langle r_1|$, and $\emph{L}_{4}=\sqrt{\gamma/2}|e_2\rangle\langle r_2|$. Thus the evolution of the system is now governed by the Markovian master equation
\begin{equation}\label{MY}
\partial_{t}\rho=-i[H_{I}^{(2)},\rho]+\sum_{i=1}^{4}\textit{L}_{i}\rho\textit{L}_{i}^{\dag}
-\frac{1}{2}\{\textit{L}_{i}^{\dag}{\textit{L}}_{i},\rho\}.
\end{equation}

The excited states $|er\rangle$, $|re\rangle$ and $|rr\rangle$ can be further adiabatically eliminated when the excited states are not initially populated. Under the second-order perturbation theory, the dynamics is given by the effective operators
\begin{equation}\label{H22}
H_{\textmd{eff}}=-\frac{1}{2}[V_{-}H_{NH}^{-1}V_{+}+V_{-}(H_{NH}^{-1})^{\dag}V_{+}],
\end{equation}
and
\begin{equation}
\emph{L}_{\textmd{eff}}^k=\emph{L}_{k}H_{NH}^{-1}V_{+},
\end{equation}
where $H_{NH}=H_{0}-\frac{i}{2}\sum_{j}\emph{L}_{j}^{\dag}\emph{L}_{j}$. Within the subspace of consideration, the effective Hamiltonian and master equation can be obtained as
\begin{figure}
\centering\scalebox{0.32}{\includegraphics{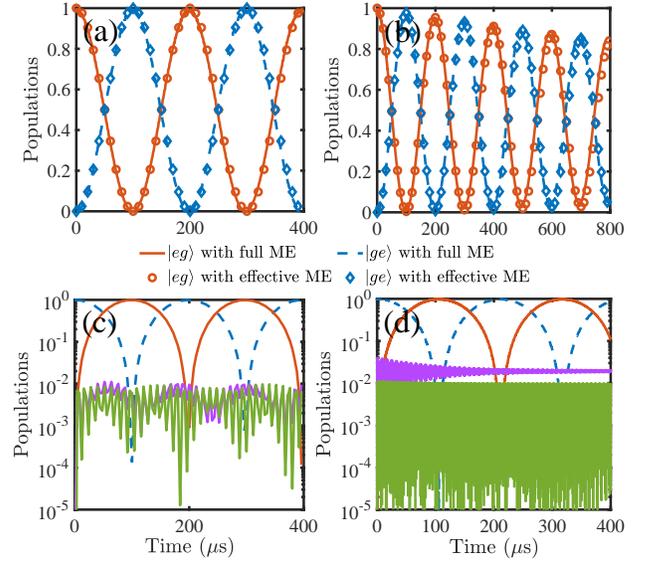}}
\caption{\label{tw1}(a) Populations of the two-atom states $|eg\rangle$ (solid) and $|ge\rangle$ (dash) governed by the full Hamiltonian of Eq.~(\ref{H1}) and the effective Hamiltonian of Eq.~(\ref{H23}), respectively. (b) Population evolutions dominated by the full master equation Eq.~(\ref{MY}) and effective master equation Eq.~(\ref{MEf}) with $\gamma=2\pi\times0.005~$MHz, respectively. (c) Populations of states $|eg\rangle$ and $|ge\rangle$ together with the singly (purple) and doubly (green) excited-states governed by the full Hamiltonian of Eq.~(\ref{H1}), where the ``excited-state" refers to the Rydberg state $|r\rangle$. The other parameters are taken as $\delta=\Omega_{p}=2\pi\times1~$MHz, $\Omega=0.05\Omega_{p}$, and $\Delta=300\Omega_{p}$. (d) Populations of states $|eg\rangle$ and $|ge\rangle$ together with the singly (purple) and doubly (green) excited-states under the protocol in Ref.~\cite{NewJ.Phys.13.073044} with parameters $\Delta_{s}=\Delta_{p}=50~$MHz, $\Omega_{s}=\Omega_{p}=7~$MHz and $U=92.5~$MHz.}
\end{figure}
\begin{equation}\label{H23}
H_{\textmd{eff}}^{(2)}=J_{12}(\sigma_{1}^{+}\sigma_{2}^{-}+\sigma_{1}^{-}\sigma_{2}^{+}),
\end{equation}
and
\begin{equation}\label{MEf}
\partial_{t}\rho=-i[H_{\textmd{eff}}^{(2)},\rho]+\sum_{i=1}^{4}\textit{L}_{\textmd{eff}}^{i}\rho\textit{L}_\textmd{eff}^{i\dag}-\frac{1}{2}\{\textit{L}_{\textmd{eff}}^{i\dag}{\textit{L}}_\textmd{eff}^{i},\rho\},
\end{equation}
where
\begin{equation}\label{L21}
\emph{L}_{\textmd{eff}}^{1}=\Gamma_{1}|ge\rangle\langle ge|+\Gamma_{2}|ge\rangle\langle eg|+\Gamma_{3}(|gr\rangle\langle ge|+|gr\rangle\langle eg|),
\end{equation}
\begin{equation}\label{L22}
\emph{L}_{\textmd{eff}}^{2}=\Gamma_{1}|eg\rangle\langle eg|+\Gamma_{2}|eg\rangle\langle ge|+\Gamma_{3}(|rg\rangle\langle ge|+|rg\rangle\langle eg|),
\end{equation}
\begin{equation}\label{L23}
\emph{L}_{\textmd{eff}}^{3}=\Gamma_{1}|ee\rangle\langle ge|+\Gamma_{2}|ee\rangle\langle eg|+\Gamma_{3}(|er\rangle\langle ge|+|er\rangle\langle eg|),
\end{equation}
\begin{equation}\label{L24}
\emph{L}_{\textmd{eff}}^{4}=\Gamma_{1}|ee\rangle\langle eg|+\Gamma_{2}|ee\rangle\langle ge|+\Gamma_{3}(|re\rangle\langle ge|+|re\rangle\langle eg|),
\end{equation}
in which $\sigma_{j}^{+}$ is a pseudo spin operator reading as $\sigma_{j}^{+}=|e_{j}\rangle\langle g_{j}|$, while $J_{12}=\Omega^{2}\Omega_{p}^{2}/(\delta^{3}-2\delta\Omega_{p}^{2})$ describes the effective coupling between ground states, $\Gamma_{1}=i\Omega\chi(\gamma^{2}-3i\gamma\delta-2\delta^{2}+2\Omega_{p}^{2})/(\gamma-2i\delta)$, $\Gamma_{2}=-2i\Omega\Omega_{p}^{2}\chi/(\gamma-2i\delta)$, and $\Gamma_{3}=\Omega\Omega_{p}\chi$ are effective spontaneous emission rates with $\chi=\sqrt{2\gamma}/(\gamma^{2}-3i\gamma\delta-2\delta^{2}+4\Omega_{p}^{2})$. The term $[\Omega_{1}\Omega_{2}(\delta^{2}-\Omega_{p}^{2})/(\delta^{3}-2\delta\Omega_{p}^{2})+\delta](|eg\rangle\langle eg|+|ge\rangle\langle ge|)$ appearing in the effective Hamiltonian has been disregarded because it only acts as a unit operator in the subspace we consider.

\subsection{Numerical simulations}
The resulting population oscillation between states $|eg\rangle$ and $|ge\rangle$ can be clearly seen in Fig.~\ref{tw1}(a) without considering the dissipative parts under parameters $\delta=\Omega_{p}=2\pi\times1~$MHz, $\Omega=0.05\Omega_{p}$, and $\Delta=300\Omega_{p}$. The evolution governed by the effective Hamiltonian of Eq.~(\ref{H23}) is well consistent with the full one of Eq.~(\ref{H1}).
Fig.~\ref{tw1}(b) shows the evolution dominated by the full master equation Eq.~(\ref{MY}) and effective master equation Eq.~(\ref{MEf}) with $\gamma=2\pi\times0.005~$MHz, respectively. The dynamics are identical to each other illustrating that the system can still be well described by a two-level form.
\begin{figure}
\centering\scalebox{0.32}{\includegraphics{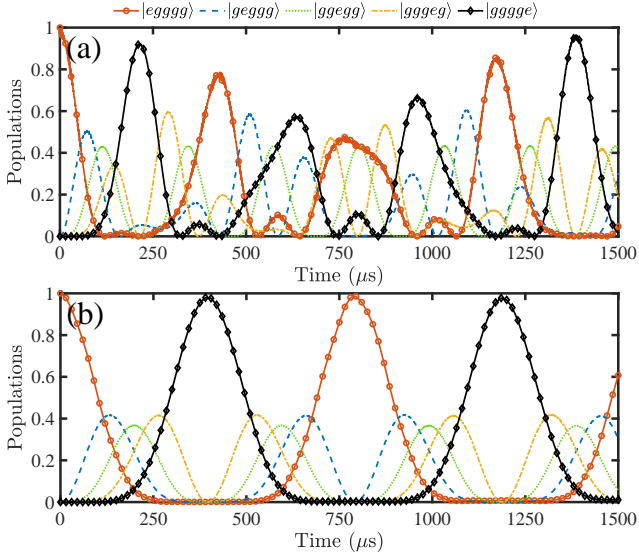}}
\caption{\label{Five}Transport dynamics of the multi-atomic model with $N=5$ without considering the spontaneous emission. (a) Populations of ground states with same separation and coupling strength governed by full Hamiltonian of Eq.~(\ref{H1})  with $\Omega_{j}=0.05\Omega_{p}$. (b) The evolution of ground states governed by full Hamiltonian under the perfect transmission condition $\Omega_{1}=\Omega_{5}=\Omega$, $\Omega_{2}=\Omega_{4}=2\Omega$, $\Omega_{3}=\sqrt{3/2}\Omega$ with $\Omega=0.025\Omega_{p}$, while the other parameters are taken as $\Omega_{p}=2\pi\times1~$MHz, $\delta=\Omega_{p}$, and ${\cal U}_{i,i+1}=\Delta=2\pi\times300~$MHz.}
\end{figure}
To be more realistic, we introduce an uncoupled state $|\alpha\rangle$ to represent the leakage levels outside qubit basis $\{|g\rangle, |e\rangle\}$, the decay term now reads as
\begin{equation}\label{1L}
{\mathcal L}=\sum_{n=1}^{2}\sum_{j=g,e,\alpha}{\it L}_{j}^{(n)}\rho{\it L}_{j}^{(n)\dagger}-\frac{1}{2}{\it L}_{j}^{(n)\dagger}{\it L}_{j}^{(n)}\rho-\frac{1}{2}\rho{\it L}_{j}^{(n)\dagger}{\it L}_{j}^{(n)},
\end{equation}
with ${\it L}_{j}^{(n)}=\sqrt{b_{jr}\gamma_r}|j_n\rangle\langle r_n|$, where $\gamma_r$ is the decay rate of the Rydberg state and $b_{jr}$ denotes the branching ratios to lower level $|j\rangle$.
The transmission efficiency corresponding to the first Rabi oscillation can be denoted as $T=\textmd{Tr}[\rho(t)\textmd{I}_{1}\otimes|e_{2}\rangle\langle e_{2}|]$, where $t=\delta\pi/2\Omega^{2}$, $\textmd{I}$ is the unit operator. Without dissipation, the excitation can be transported to the second atom perfectly, and hence $T=1$, for an ideal Heisenberg $XX$ spin-chain model. In the presence of the dissipation, we use $T^{\lambda}$ to denote the transmission efficiency with different branching ratio $\lambda=b_{gr}+b_{er}$.
In the hyperfine structure, by assuming that the dissipation rates from the Rydberg state to any ground state are equal to each other, we have $b_{gr}=1/8, b_{er}=1/8$ and $b_{\alpha r}=3/4$ \cite{PhysRevA.101.062309}.
The corresponding transmission efficiency is calculated as $T^{0.25}=0.9736$. With the pessimistic approximation that $b_{gr}+b_{er}=0$, $b_{\alpha r}=1$, we still have $T^{0}=0.9716$, which proves that the spontaneous radiation out of space has little effect on the system.
In Fig.~\ref{tw1}(c) and \ref{tw1}(d), we further compare our scheme with the method provided in Ref.~\cite{NewJ.Phys.13.073044}, where the populations of singly (purple) and doubly (green) excited states (the ``excited-state" refers to the Rydberg state $|r\rangle$) are simulated by the full Hamiltonian of Eq.~(\ref{H1}) and the Hamiltonian of Eq.~(4) in the Ref.~\cite{NewJ.Phys.13.073044}, respectively. Under the condition of realizing the same Rabi oscillation period between $|eg\rangle$ and $|ge\rangle$,
it can be clearly seen that our scheme has better effect on inhibiting atomic excitation.
In addition, this diatomic model can also be used to implement the $\sqrt{\textmd{SWAP}}$ gate which together with single-qubit rotations form a set of universal gates for quantum computation \cite{PhysRevA.70.062302,science.1116955,nature06011,Zhang_2011,Shao_2014} (please see Appendix A for details).

Since the long-range vdW interaction between the next nearest neighbour atoms is too weak to fulfill the condition $\Delta={\cal U}_{i,i+2}$, these terms can be neglected as high-frequency oscillating terms with detuning $\Delta-{\cal U}_{i,i+2}$. The effective Hamiltonian for arbitrary $N$ particles reduces to
\begin{equation}\label{Ar1}
H_{\textmd{eff}}^{(N)}=\sum_{j=1}^{N-1}J_{j,j+1}(\sigma_j^{+}\sigma_{j+1}^{-}+\sigma_j^{-}\sigma_{j+1}^{+}),
\end{equation}
where $J_{j,j+1}=\Omega_j\Omega_{j+1}\Omega_p^2/(\delta^{3}-2\delta\Omega_p^2)$. Note that when $\Omega_{p}$ and $\delta$ hold the same magnitude, the form of coupling strength between ground states is simplified as $-\Omega_{j}\Omega_{j+1}/\delta$ which is only related to the properties of the weak driving fields. As $J_{j,j+1}=J$, Eq.~(\ref{Ar1}) is equivalent to a Heisenberg $XX$ spin chain restricted in the single-excitation manifold. Fig.~\ref{Five}(a) depicts the spin-chain dynamics of five particles governed by the full Hamiltonian of Eq.~(\ref{H1}) from the initial state $|egggg\rangle$. The corresponding parameters are taken as $\delta=\Omega_{p}=2\pi\times1~$MHz, $\Delta=2\pi\times300~$MHz, and $\Omega_{j}=0.05\Omega_{p}$. Meanwhile, the perfect quantum state transfer can be also achieved by tuning the Rabi frequencies $\Omega_{j}$ to meet the condition $J_{j,j+1}=J\sqrt{j(N-j)}$ \cite{PhysRevLett.92.187902}. For five particles case, the corresponding parameters can be selected as $\Omega_{1}=\Omega_{5}=\Omega$, $\Omega_{2}=\Omega_{4}=2\Omega$, and $\Omega_{3}=\sqrt{3/2}\Omega$, where $\Omega=0.025\Omega_{p}$. The populations of single-excited states under the full Hamiltonian of Eq.~(\ref{H1}) are shown in Fig.~\ref{Five}(b).

\begin{figure}\scalebox{0.32}{\includegraphics{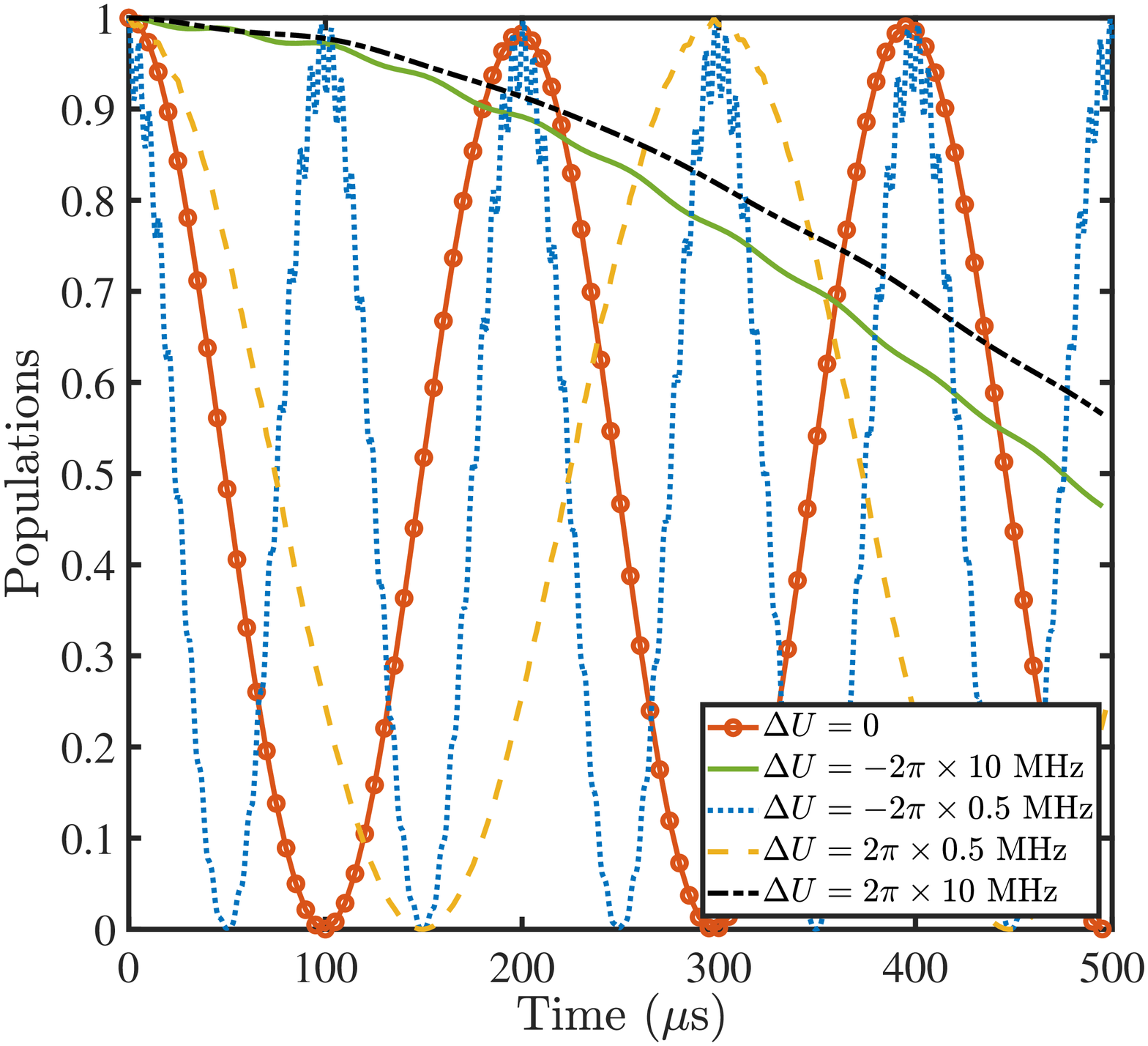}}
\caption{\label{Error}The mismatch of the unconventional Rydberg pumping condition governed by full Hamiltonian of Eq.~(\ref{H1}). The parameters are taken as $\delta=\Omega_{p}=2\pi\times1~$MHz, $\Delta=2\pi\times300~$MHz, $\Omega=0.05\Omega_{p}$, and $\Delta U={\cal U}_{j,j+1}-\Delta$.}
\end{figure}

Experimentally, there may be a systematic error in the position of atoms destroying the condition ${\cal U}_{j,j+1}=\Delta$. In order to investigate the influence of this factor, we introduce the degree of deviation $\Delta U={\cal U}_{j,j+1}-\Delta$. To ensure that the near-resonance terms kept before are still dominant, we assume that $\Delta U$ is not very large.
After calculation by effective operator method, the effective Hamiltonian keeps the same form as Eq.~(\ref{H23}) but with an updated coupling strength related to $\Delta U$
\begin{equation}\label{ED4}
J_{12}=\frac{\Omega^{2}\Omega_{p}^{2}}{\delta^{3}-2\delta\Omega_p^{2}-\delta^{2}\Delta U}.
\end{equation}
Setting $\delta=\Omega_{p}=2\pi\times1~$MHz, $\Delta=2\pi\times300~$MHz and $\Omega=0.05\Omega_{p}$, the dynamical evolution modeled by different $\Delta U$ is  shown in Fig.~\ref{Error}. With such parameters, we have $\delta^{3}-2\delta\Omega_{p}^{2}<0$. According to Eq.~(\ref{ED4}),
when $\Delta U>0$, the increase value of $|\Delta U|$ will lead to the decrease of coupling coefficient $J_{12}$ and the extension of evolution period. When $\Delta U<0$, the evolution period is shortened under the condition $0<|\Delta U|<(1+2\delta\Omega_{p}^{2}-\delta^{3})/\delta^{2}$ and extended  under the condition $|\Delta U|>(1+2\delta\Omega_{p}^{2}-\delta^{3})/\delta^{2}$. Note that $\Delta U=-(1+2\delta\Omega_{p}^{2}-\delta^{3})/\delta^{2}$ is a singularity of Eq.~(\ref{ED4}), which will destroy the condition of
second-order perturbation and should be avoided when considering the actual physical parameters.
On the whole, in the presence of a small deviation, the above derivation process is still valid and the effective coupling strength $J_{j,j+1}$ becomes a function of $\Delta U$ which will only change the evolution cycle of the system but will not invalidate the scheme.
Here, we only consider the case where the atomic position is fixed for simplicity, the random fluctuation of vdW interaction caused by the atomic vibration will be further discussed in section \ref{5b}.

\section{Topological spin model}\label{III}
\begin{figure}
\centering
\includegraphics[scale=0.2]{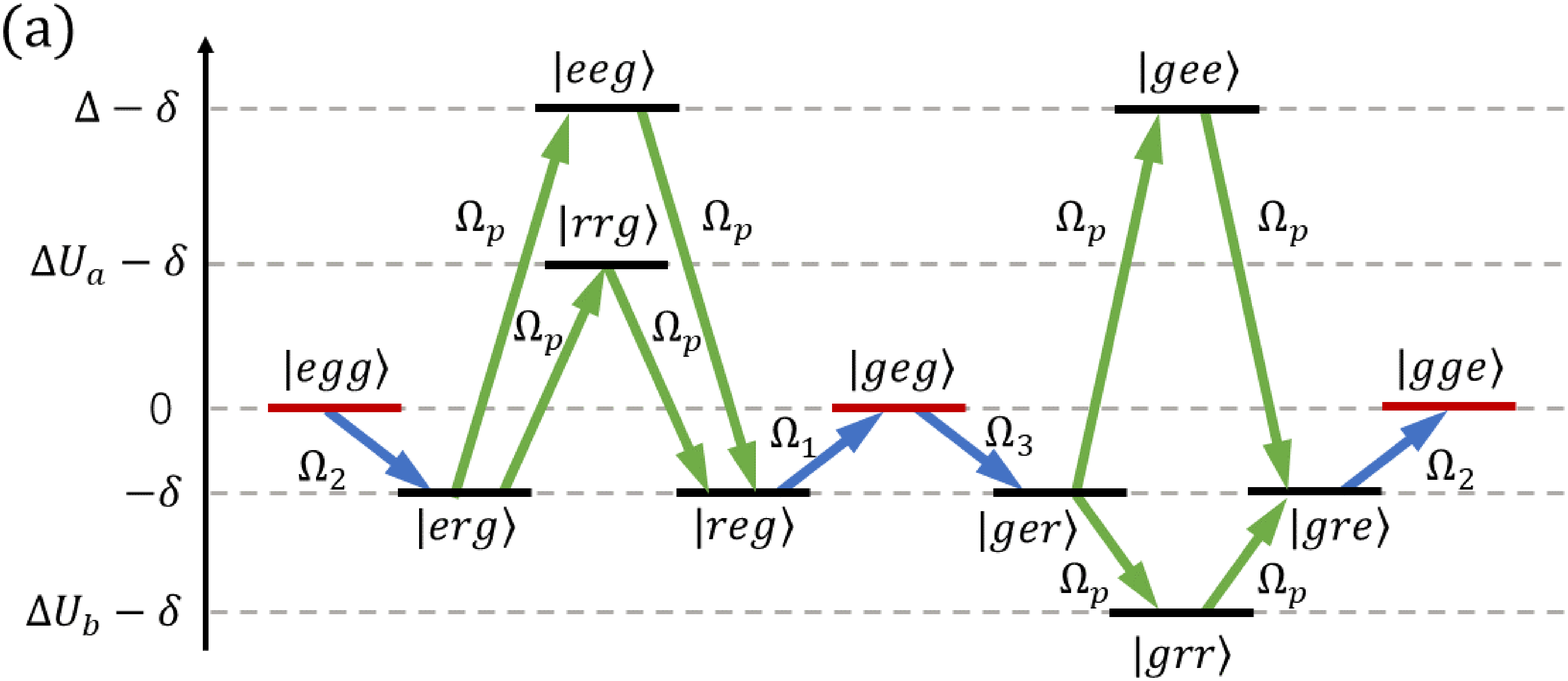}
\hspace{1in}
\includegraphics[scale=0.25]{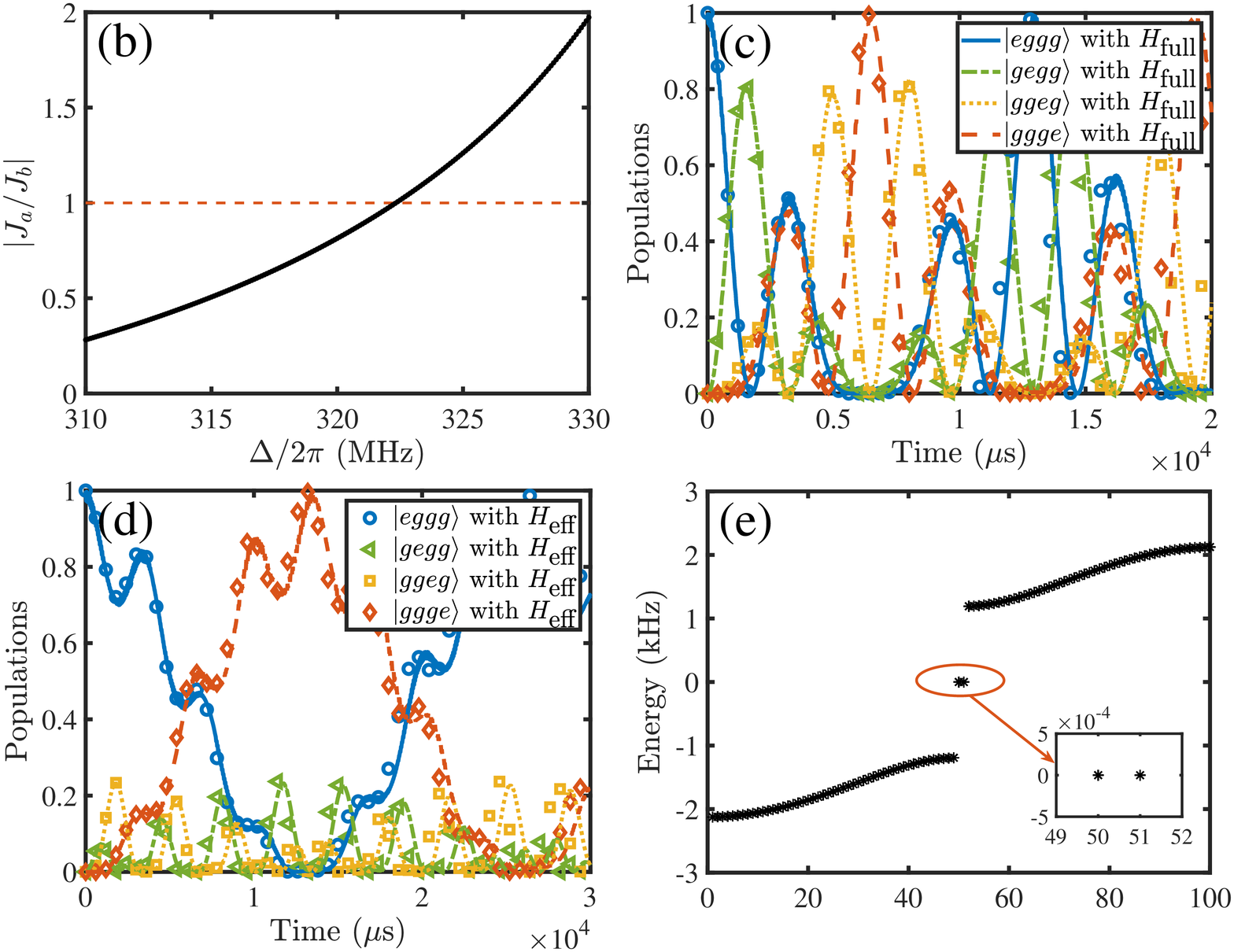}
\caption{\label{Tople}(a) The effective coupling processes for topological spin model. (b) The ratio of $J_{a}$ to $J_{b}$ is shown as a function of the detuning $\Delta$. (c) Populations of ground states $|eggg\rangle$ (solid), $|gegg\rangle$ (dot-dash), $|ggeg\rangle$ (dotted) and $|ggge\rangle$ (dash) governed by full and effective Hamiltonian shown as Eqs.~(\ref{H1}) and (\ref{DE1}) with $\Delta=2\pi\times330~$MHz. (d) The population evolutions of ground states with $\Delta=2\pi\times310~$MHz. (e) The energy spectrum with topological phase with $N=100$, where $\Delta=2\pi\times310~$MHz. The other parameters are taken as $\delta=\Omega_{p}=2\pi\times1~$MHz, $\Omega_{j}=0.05\Omega_{p}$, $r_{2i-1,2i}=4~\mu$m, and $r_{2i,2i+1}=4.1~\mu$m.}
\end{figure}
Inspired by the inherent adjustable coupling of the system, we show that the following 1D SSH model can be constructed \cite{PhysRevLett.42.1698,PhysRevLett.62.2747,JK2016A,PhysRevE.102.012101,OSTAHIE2021127030,Jiang2020,science.aav9105}
\begin{equation}\label{DE1}
H_{\textmd{ssh}}=\sum_{i=1}^\frac{N}{2}J_{a}\sigma_{2i-1}^{+}\sigma_{2i}^-+\sum_{i=1}^{\frac{N}{2}-1}J_{b}\sigma_{2i}^{+}\sigma_{2i+1}^-+{\rm H.c.},
\end{equation}
with regard to even number of particles, where $J_a$ and $J_b$ represent real intra- and inter-unit-cell
hopping coefficients, respectively.
Different from the previous scheme (i.e. $U=\Delta$), we here take the deviation of the unconventional Rydberg pumping condition $\Delta U$ as a control parameter to achieve our goal.
When ${\cal U}_{j,j+1}$ is significantly different from $\Delta$ ($\Delta U$ is relatively large), the extra coupling induced by doubly ``excited" states with two atoms in state $|e\rangle$ while others in $|g\rangle$ should also be taken into account. For the simplest system composed of three particles with non-identical coupling, the transition paths are shown in Fig.~\ref{Tople}(a).
\begin{figure}\scalebox{0.32}{\includegraphics{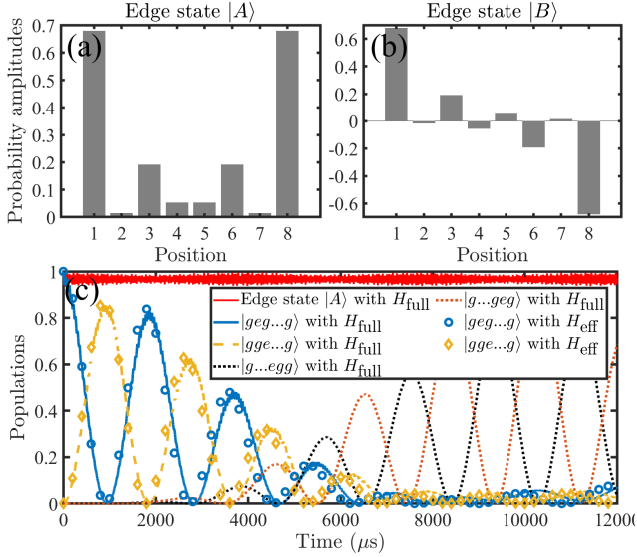}}
\caption{\label{Eight}The transport dynamics and the edge states of the system with $N=8$. (a) and (b) correspond to the probability amplitudes of two edge states governed by the effective Hamiltonian of Eq.~(\ref{DE1}). (c) Populations of the edge state $|A\rangle$ governed by the full Hamiltonian of Eq.~(\ref{H1}) (solid line) and the populations of states $|geg...g\rangle$ and $|gge...g\rangle$ governed by the full (dash and dotted) and effective (diamond and circle) Hamiltonian initially excited at second particle, respectively. The parameters are taken as $\delta=\Omega_{p}=2\pi\times1~$MHz, $\Omega=0.05\Omega_{p}$, $\Delta=2\pi\times310~$MHz, $r_{2i-1,2i}=4~\mu$m, and $r_{2i,2i+1}=4.1~\mu$m.}
\end{figure}
Thus the form of $H_{NH}$ and $V_{+}$ should be rewritten as
\begin{eqnarray}
H_{NH}&=&(\Delta U_{a}-\delta)|rrg\rangle\langle rrg|-\delta(|erg\rangle\langle erg|+|reg\rangle\langle reg|\nonumber\\&&
+|ger\rangle\langle ger|+|gre\rangle\langle gre|)+(\Delta U_{b}-\delta)|grr\rangle\langle grr|\nonumber\\&&
+(\Delta-\delta)(|eeg\rangle\langle eeg|+|gee\rangle\langle gee|)
\nonumber\\&&+\Omega_{p}(|ger\rangle\langle grr|+|gre\rangle\langle grr|+|ger\rangle\langle gee|\nonumber\\&&+|gre\rangle\langle gee|+|erg\rangle\langle rrg|+|reg\rangle\langle rrg|\nonumber\\&&+|erg\rangle\langle eeg|+|reg\rangle\langle eeg|+{\rm H.c.)},
\end{eqnarray}
and
\begin{eqnarray}
V_{+}&=&\Omega_2|erg\rangle\langle egg|+\Omega_1|reg\rangle\langle geg|+\Omega_3|ger\rangle\langle geg|\nonumber\\&&+\Omega_2|gre\rangle\langle gge|,
\end{eqnarray}
where $\Delta U_{k}={\cal U}_{k}-\Delta$, $k=a,b$, ${\cal U}_{a}\propto1/r_{2i-1,2i}^{6}$ and ${\cal U}_{b}\propto1/r_{2i,2i+1}^{6}$. Assuming $\Omega_{j}=\Omega$ and applying the effective operator method again, the Rydberg states can be adiabatically eliminated and the effective coupling strength between ground states can be obtained as
\begin{equation}\label{du1}
J_{k}=-\frac{\Omega^{2}\Omega_{p}^{2}({\cal U}_{k}-2\delta)}{\delta^{4}-{\cal U}_{k}\delta^{3}-\eta_{k}\delta^{2}+2{\cal U}_{k}\Omega_{p}^{2}\delta},
\end{equation}
where $\eta_{k}=4\Omega_{p}^{2}+\Delta^{2}-{\cal U}_{k}\Delta$. If the two distances between atoms are set to be $r_{2i-1,2i}=4\mu$m and $r_{2i,2i+1}=4.1\mu$m, the corresponding vdW interactions are  ${\cal U}_{a}\simeq2\pi\times346~$MHz and ${\cal U}_{b}\simeq2\pi\times300~$MHz. Therefore, after setting other parameters as $\delta=\Omega_{p}=2\pi\times1~$MHz and $\Omega=0.05\Omega_{p}$, the effective coupling strength $J_{k}$ becomes a single valued function of $\Delta$.  As illustrated in Fig.~\ref{Tople}(b), the ratio $|J_{a}/J_{b}|$ is related to topological phase. When $\Delta\in2\pi\times[310,323)$~MHz, the system corresponds to a nontrivial topological phase for $|J_{a}/J_{b}|<1$, where an additional state localized at the boundaries around zero energy can be observed. When $\Delta\in2\pi\times[323,330]$~MHz, the system corresponds to a trivial phase for $|J_{a}/J_{b}|>1$ with two discrete energy bands.
\begin{figure}
\centering
\includegraphics[scale=0.2]{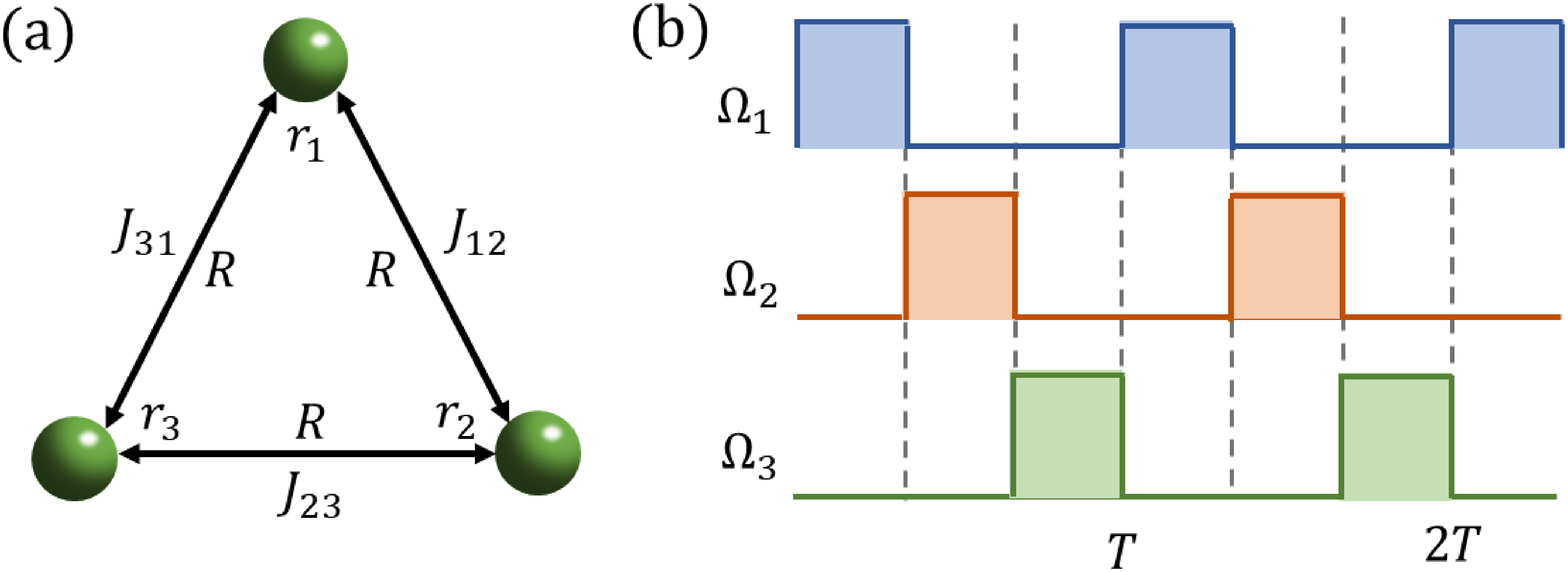}
\hspace{1in}
\includegraphics[scale=0.2]{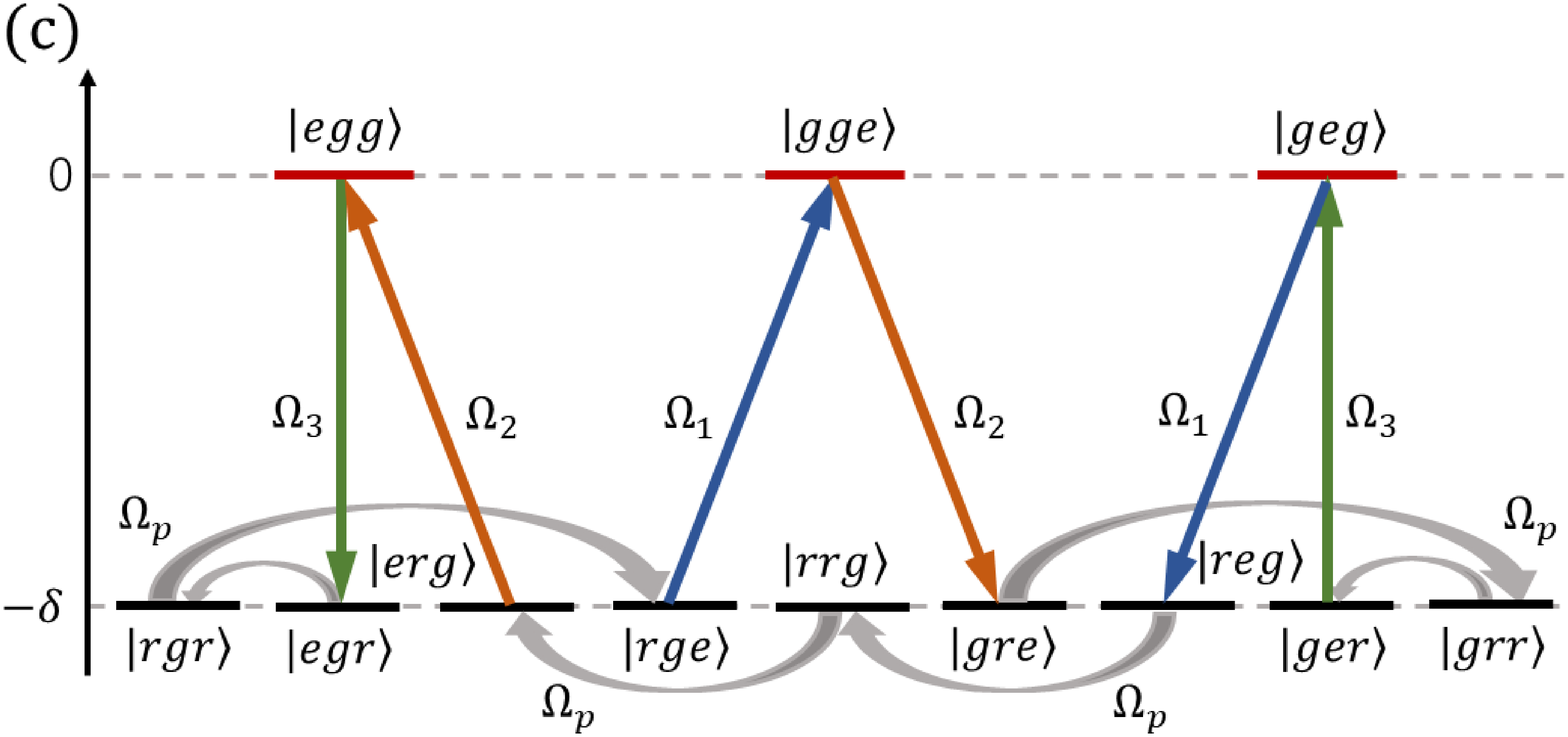}
\caption{\label{current1}(a) Schematic representations of Rydberg atoms arranged as equilateral triangle form, $J_{j k}$ describes the effective coupling between the $j$th and $k$th atom. (b) Periodic modulated pulses for realizing the chiral motion of atomic excitation. (c) The effective coupling processes have been chosen for realizing chiral motion.}
\end{figure}
\begin{figure*}\scalebox{0.35}{\includegraphics{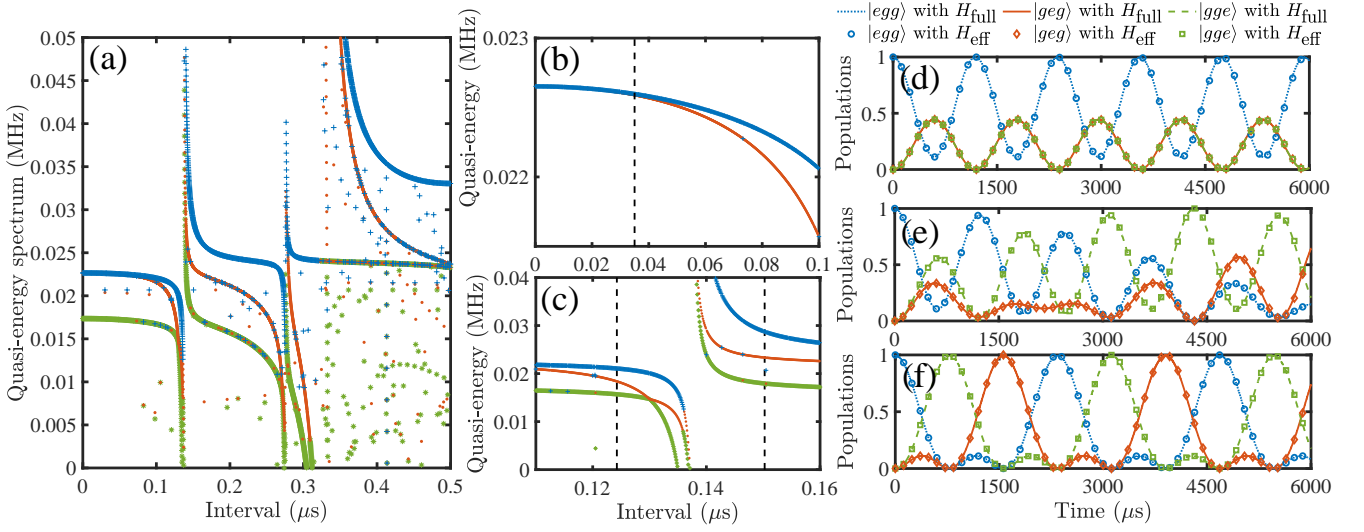}}
\caption{\label{Curr_Sp}(a) The quasi-energy spectrum of the effective Hamiltonian $H_{\textmd{eff}}=i\textmd{ln}(e^{-iH_{3}\tau}e^{-iH_{2}\tau}e^{-iH_{1}\tau})/3\tau$ under various time intervals $\tau$. (b) and (c) are the enlarged view of the quasi-energy spectrum, which respectively show the energy splitting process and the position the chiral motion generated. (d)-(f) show the populations of the ground states under different time intervals respectively corresponding to the quasi-energy spectrum with two degenerated eigenenergy, arbitrary split three eigenenergy, and the eigenenergy with same separations, where $\tau=0.01\mu$s, $0.1\mu$s, and $0.12425\mu$s. The parameters are taken as $\delta=\Omega_{p}=2\pi\times1~$MHz, $\Omega=0.05\Omega_{p}$, and $\Delta=2\pi\times300~$MHz.}
\end{figure*}
\begin{figure*}
\centering\scalebox{0.35}{\includegraphics{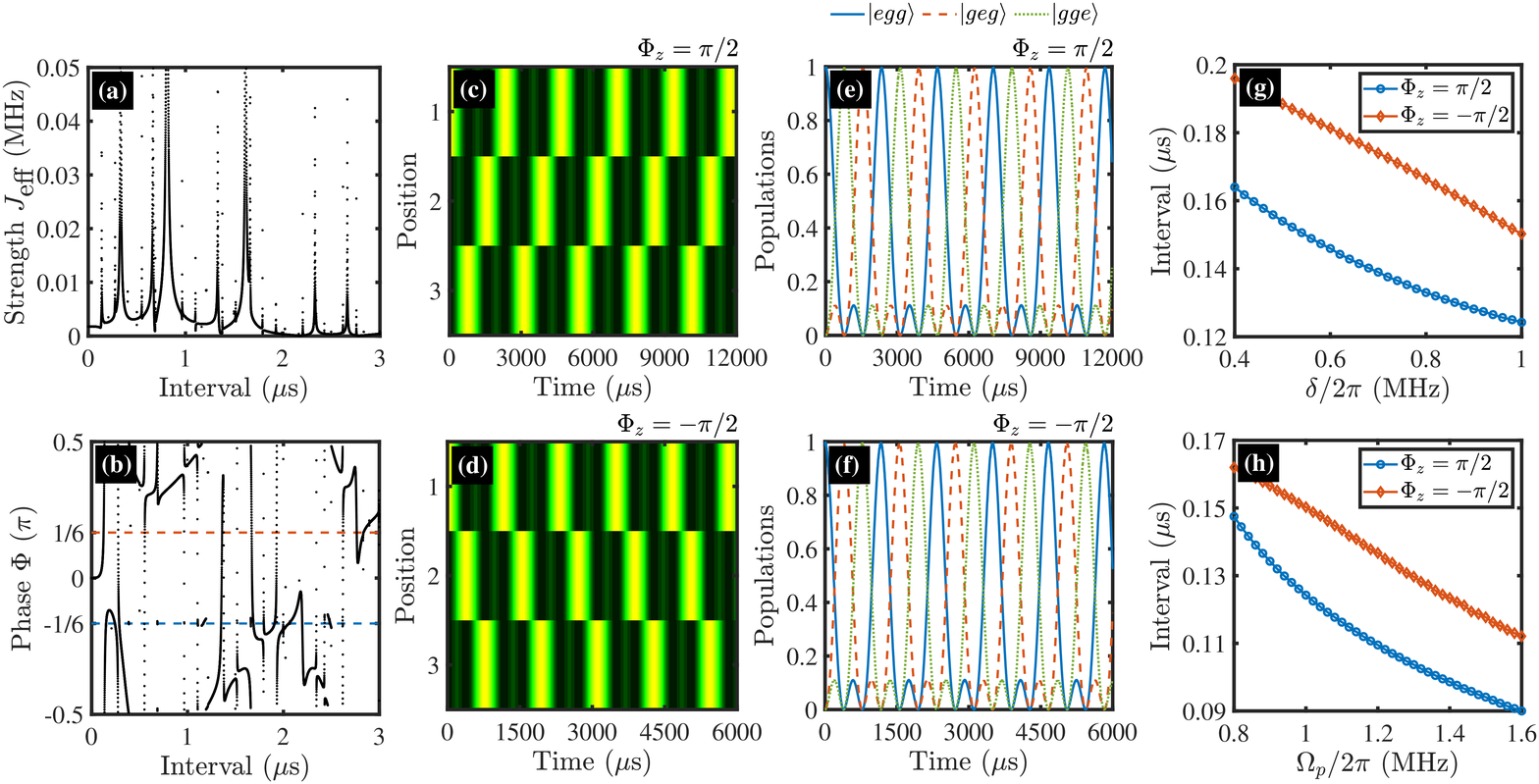}}
\caption{\label{current}The chiral motion of atomic excitation under the equilateral triangle structure. (a) and (b) respectively show the average value of the effective coupling strength and induced phases under different time intervals. (c) and (d) intuitively represent the transfer of the ground state between three positions. (e) and (f) further show the populations of ground states as a function of time measured for values of $\Phi_{z}=\pi/2, -\pi/2$ while the time interval is respectively taken as $0.12425~\mu$s and $0.15025~\mu$s. The parameters are setting as $\delta=\Omega_{p}=2\pi\times1~$MHz, $\Omega_{j}=0.05\Omega_{p}$, and $\Delta=2\pi\times300~$MHz. (g) and (h) exhibit the time interval $\tilde{\tau}$ as functions of parameters $\delta$ and $\Omega_{p}$, respectively.}
\end{figure*}
So the system can be modulated from the nontrivial topological phase to trivial topological phase by regulating the detuning $\Delta$ from $2\pi\times310~$MHz to $2\pi\times330~$MHz within the current parameter setting range.
Fig.~\ref{Tople}(c) and \ref{Tople}(d) respectively show the evolution of ground states of $N=4$ with $\Delta=2\pi\times330~$MHz and $\Delta=2\pi\times310~$MHz. The evolution governed by the effective Hamiltonian of Eq.~(\ref{DE1}) is well consistent with the full Hamiltonian of Eq.~(\ref{H1}) which proves that the approximation is valid.
In Fig.~\ref{Tople}(e), we characterize the energy spectrum of a multi-particle ($N=100$) SSH model in the case of single excitation with $\Delta=2\pi\times310~$MHz. The gap between the edge state and bulk is about $1.2~$kHz which corresponds to a nontrivial topological phase with two zero-energy edge states.

Fig.~\ref{Eight} further reveals the edge states and the transport dynamics of the system with eight particles. Fig.~\ref{Eight}(a) and \ref{Eight}(b) are corresponding to the probability amplitudes of two edge states governed by the effective Hamiltonian, which are mainly distributed on two sides. Fig.~\ref{Eight}(c) shows the evolution results for two different initial conditions, the edge state $|A\rangle$ and $|gegg...\rangle$. The edge state $|A\rangle$ does not transfer to other states and oscillates around $1$, but the population of state $|geg...g\rangle$ will transfer to the intermediate particles except the two sides. These behaviors characterize as the topological structure. We can also see from Fig.~\ref{Eight} that the dynamics described by the effective Hamiltonian (marked with circle and diamond) is the same as the full Hamiltonian.
Therefore, using the ground state of Rydberg atom can construct an effective SSH model, and help to further provides an alternative way for realization of quantum state transmission based on topological model \cite{PhysRevA.98.012331,PhysRevResearch.2.033475}.

\section{Chiral motion of atomic excitation}\label{IV}
Compared with the tight-binding model with open boundary condition, the tight-binding model with periodic boundary condition can exhibit more abundant physical properties. For three particles arranged in an equilateral triangle shown in Fig.~\ref{current1}(a), the following form of Hamiltonian can exist under the induction of gauge field
\begin{equation}\label{Ec1}
H_{\Phi_{z}}=-J_{c}(e^{i\phi_{12}}\sigma_{1}^{+}\sigma_{2}^{-}+e^{i\phi_{23}}\sigma_{2}^{+}\sigma_{3}^{-}+e^{i\phi_{31}}\sigma_{3}^{+}\sigma_{1}^{-})+{\rm H.c.},
\end{equation}
where $J_{c}$ is a positive real number and $\Phi_{z}=\phi_{12}+\phi_{23}+\phi_{31}$ can be seen as a synthetic flux behaving similarly to physical magnetic flux. When $\Phi_{z}=\pi/2$, the atomic excitation $|e\rangle$ propagates in the counter clockwise direction $1\rightarrow3\rightarrow2\rightarrow1$. When $\Phi_{z}=-\pi/2$, the direction of transmission reverses. Ever the synthetic flux $\Phi_{z}\neq0$, the dynamics of the system breaks the time-reversal symmetry known as a chiral motion \cite{nphys3930,PhysRevX.10.021031}.

In our scheme, this chiral motion can be simulated by periodic modulation under the floquet theorem \cite{doi:10.1080/00018732.2015.1055918,PhysRevA.75.063424,PhysRevLett.108.156602,PhysRevLett.91.057401,PhysRevB.89.184516,
PhysRevB.91.054517}, and the phase of the
hopping amplitude between adjacent sites can be induced through the noncommutativity between Hamiltonian. To be specific, the piecewise constant Hamiltonian is shown in the form of
\begin{equation}\label{Ecc2}
 H(t)=\left\{
              \begin{array}{lr}
              H_{1},\ \ t\in[0, T/3)\\
              H_{2},\ \ t\in[T/3, 2T/3)\\
              H_{3},\ \ t\in[2T/3, T)\\
              \end{array}
\right.
\end{equation}
where
\begin{eqnarray}\label{C1}
H_{i}&=&\Omega_{i}|r_{i}\rangle\langle g_i|+\sum_{j=1}^{3}\Omega_{p}|r_{j}\rangle\langle e_j|+{\rm H.c.}
+\delta|g_{j}\rangle\langle g_j|\nonumber\\&&+\Delta|e_{j}\rangle\langle e_j|+\sum_{j<k}{\cal U}_{j k}|r_jr_k\rangle\langle r_jr_k|.
\end{eqnarray}
In this case, the evolution period of the system has been set as $T$, it contains three processes and each one is described by $H_{i}$ corresponding to the evolution time $\tau=T/3$. This can be achieved by switching on and off the weak fields coupled to the transition between states $|g\rangle$ and $|r\rangle$ of atom in sequence. Considering the effect of all 27 states on the system dynamics during alternation, the effective Hamiltonian is presented as $H_{\textmd{eff}}=i\textmd{ln}(e^{-iH_{3}\tau}e^{-iH_{2}\tau}e^{-iH_{1}\tau})/T$ \cite{2000Quantum}.
The quasi-energy spectrum of this effective Hamiltonian under various time intervals is displayed in Fig.~\ref{Curr_Sp}(a), keeping only the eignenergies in the ground-state manifold constructed by $|egg\rangle$, $|geg\rangle$, and $|gge\rangle$ for the sake of clarity. When the time interval $\tau<0.035~\mu$s, there exists a double-degenerated quasi-energy spectrum and the effective Hamiltonian can be well described by the Trotter product formula $\lim_{N\rightarrow\infty}\{e^{-iH_{3}T/3N}e^{-iH_{2}T/3N}e^{-iH_{1}T/3N}\}^{N}=e^{-i(H_{1}+H_{2}+H_{3})T/3}$ \cite{Simon1980Methods,Engel2001},
Then we have $H_{\textmd{eff}}=\sum_{j=1}^3J_{j,j+1}'(\sigma_{j}^{+}\sigma_{j+1}^{-}+\sigma_{j}^{-}\sigma_{j+1}^{+})$, where $J_{j,j+1}'=\Omega_{j}\Omega_{j+1}\Omega_{p}^2/9(\delta^3-2\delta\Omega_{p}^2)$, ($\Omega_{4}=\Omega_{1}$). This condition is identical to the population evolution of ground states with $\tau=0.01~\mu$s shown in Fig.~\ref{Curr_Sp}(d).
Due to the influence of the hopping phase, the degeneracy is removed as $\tau$ gets longer and the population evolution are shown in Fig.~\ref{Curr_Sp}(e) with $\tau=0.1~\mu$s.
When the difference between the eigenenergies is equal as indicated in Fig.~\ref{Curr_Sp}(c), the phase induced by the alternate evolution are just $\pm\pi/2$ resulting in a directional chiral motion of atomic excitation, as shown in Fig.~\ref{Curr_Sp}(f) for $\tau=0.12425~\mu$s.

A combination of numerical and analytical methods can be employed to determine the specific value of $\Phi_{12(23,31)}$ corresponding to various time intervals. According to Eq.~(\ref{C1}), under the condition
${\cal U}_{j,j+1}=\Delta\gg\{\Omega_{p},\delta\}\gg\Omega_{i}$, we can neglect the high-frequency oscillating terms and the dynamics of the system are mainly restricted in the subspace constructed by $|egg\rangle$, $|egr\rangle$, $|rgr\rangle$, $|rge\rangle$, $|gge\rangle$, $|gre\rangle$, $|grr\rangle$, $|ger\rangle$, $|geg\rangle$, $|reg\rangle$, $|rrg\rangle$, and $|erg\rangle$, as shown in Fig.~\ref{current1}(c). In order to simplify computational space, we complete the following calculations in the subspace composed of above 12 states. Once interval $\tau$ is given, the effective Hamiltonian can be numerically obtained via the second order perturbation theory. Keeping the convergent results and discard the divergent results, we have the effective Hamiltonian form as
\begin{eqnarray}\label{cu}
H_{\textmd{eff}}&=&-J_{\textmd{eff}}^{12}e^{i\Phi_{12}}|egg\rangle\langle geg|-J_{\textmd{eff}}^{23}e^{i\Phi_{23}}|geg\rangle\langle gge|\nonumber\\&&-J_{\textmd{eff}}^{31}e^{i\Phi_{31}}|gge\rangle\langle egg|+{\rm H.c.},
\end{eqnarray}
corresponding to a certain time interval $\tau$.
The average values of $J_{\textmd{eff}}$ [$J_{\textmd{eff}}=1/3(J_{\textmd{eff}}^{12}+J_{\textmd{eff}}^{23}+J_{\textmd{eff}}^{31})$] and $\Phi$ [$\Phi=1/3(\Phi_{12}+\Phi_{23}+\Phi_{31})$] versus different time intervals are displayed in Fig.~\ref{current}(a) and \ref{current}(b), respectively. From which we can read the specific value of $J_{\textmd{eff}}$ and $\Phi$ for any $\tau$. For example, when $\tau=0.01\mu$s, we have $J_{\textmd{eff}}\simeq1.763~$kHz, $\Phi\simeq0$, when $\tau=0.1~\mu$s, we have $J_{\textmd{eff}}\simeq1.7~$kHz, $\Phi\simeq0.0263\pi$, and when $\tau=0.12425~\mu$s, we have $J_{\textmd{eff}}\simeq1.55~$kHz, $\Phi\simeq0.1643\pi$. The system dynamics corresponding to $\tau=\{0.01, 0.1, 0.12425\}~\mu$s governed by the effective Hamiltonian of Eq.~(\ref{cu}) with above $J_{\textmd{eff}}$ and $\Phi$ are shown in Fig.~\ref{Curr_Sp}(d)-(f). The evolutions of populations coincide with the full one of Eq.~(\ref{Ecc2}), which indicates that the effective results are reliable.
According to the continuity equation on the lattice \cite{PhysRevA.98.053812,PhysRevB.91.094502,PhysRevLett.120.157201}
\begin{equation}
\frac{d}{dt}\sigma_{j}^{z}=i[H_{\textmd{eff}},\sigma_{j}^{z}]=\triangledown_{j}I_{j}=I_{j,j-1}-I_{j,j+1},
\end{equation}
where $\sigma_{j}^{z}=|e_j\rangle\langle e_j|-|g_j\rangle\langle g_j|$, one can find that the expectation value of the current operator in the ground state of Eq.~({\ref{cu}}) for the bond $j\rightarrow j+1$ on the lattice is given by
\begin{equation}
\langle I_{j,j+1}\rangle=2iJ_{\textmd{eff}}^{j,j+1}(e^{i\Phi_{j,j+1}}\langle\sigma_{j}^{+}\sigma_{j+1}^{-}\rangle-{\rm c.c.}),
\end{equation}
 Corresponding to Fig.~\ref{Curr_Sp}(d)-(f), the ground-state current for the bond $1\rightarrow2$ are measured as $\{0, -0.0596, -0.3376\}~$kHz.

When $\Phi=\pm\pi/6$, the chiral motion holding a definite direction can be successfully obtained since the effective Hamiltonian of the system fits Eq.~(\ref{Ec1}). As can be seen from Fig.~\ref{current}(b), there are multiple time intervals can be selected, and we only choose the shortest time to discuss for convenience. Fig.~\ref{current}(c)[(e)] and \ref{current}(d)[(f)] respectively show the excited population transport between different atoms governed by full Hamiltonian of Eq.~(\ref{Ecc2}) with $\tilde{\tau}_{ac}=0.12425~\mu$s and $\tilde{\tau}_{c}=0.15025~\mu$s. Note that $\tilde{\tau}$ represents the time interval at which $\Phi=\pm\pi/6$ in later descriptions. From the atomic arrangement shown in Fig.~\ref{current1}(a), we can see that $\tilde{\tau}_{ac(c)}$ leads to the anticlockwise (clockwise) current. Thus the direction of motion can be controlled by changing $\tau$. The parameters have been taken as $\delta=\Omega_{p}=2\pi\times1~$MHz, $\Omega=0.05\Omega_{p}$, and $\Delta={\cal U}_{j,k}=2\pi\times300~$MHz, and the effective coupling strength $J_{\textmd{eff}}^{1(2)}\simeq1.55$ $(3.14)~$kHz matching $\tilde{\tau}_{ac(c)}$. Since the effective coupling strength is a function of $\Omega_{j}$, $\Omega_{p}$ and $\delta$, the time interval $\tilde{\tau}$ required to get this directional chiral motion possibly related to these parameters. We have performed numerical simulation under different parameters and the comparison shows that $\tilde{\tau}$ is closely related to $\delta$ and $\Omega_{p}$.
With parameters $\Omega=0.05\Omega_{p}$, and $\Delta={\cal U}_{j,k}=2\pi\times300~$MHz, Fig.~\ref{current}(g) and \ref{current}(h) further characterize the change of $\tilde{\tau}$ with $\delta$ for a fixed $\Omega_{p}=2\pi\times1~$MHz and with $\Omega_{p}$ for a fixed $\delta= 2\pi\times1~$MHz, respectively.
After polynomial fitting, we obtain $\tilde{\tau}$ as functions of $\delta$ and $\Omega_{p}$ presented as $\tilde{\tau}=p_{5}\delta^{4}+p_{4}\delta^{3}+p_{3}\delta^{2}+p_{2}\delta+p_{1}$ and $\tilde{\tau}=q_{5}\Omega_{p}^{4}+q_{4}\Omega_{p}^{3}+q_{3}\Omega_{p}^{2}+q_{2}\Omega_{p}+q_{1}$, where the coefficients corresponding to $\Phi_{z}=\pm\pi/2$ are shown in Table.~\ref{coe}. Therefore, according to any $\delta\in[0.8\pi,2\pi]~$MHz or $\Omega_{p}\in[1.6\pi,3.2\pi]~$MHz, the time interval $\tilde{\tau}$ can be estimated through the above functions.

\begin{table}
\centering
\caption{\label{coe}The coefficients corresponding to numerical fitting results.}
\begin{tabular}{ll}
\hline\hline
$\Phi_{z}=\pi/2$&$\Phi_{z}=-\pi/2$\\
\hline
$p_{1}=0.23278~\textmd{MHz}^{-1}$&$p_{1}=0.24661~\textmd{MHz}^{-1}$\\
$p_{2}=-4.0934\times10^{-2}~\textmd{MHz}^{-2}$&$p_{2}=-3.2661\times10^{-2}~\textmd{MHz}^{-2}$\\
$p_{3}=7.0238\times10^{-3}~\textmd{MHz}^{-3}$&$p_{3}=7.3238\times10^{-3}~\textmd{MHz}^{-3}$\\
$p_{4}=-7.3771\times10^{-4}~\textmd{MHz}^{-4}$&$p_{4}=-1.0702\times10^{-3}~\textmd{MHz}^{-4}$\\
$p_{5}=3.4895~\textmd{MHz}^{-5}$&$p_{5}=5.4603\times10^{-5}~\textmd{MHz}^{-5}$\\
\hline
$q_{1}=0.52719~\textmd{MHz}^{-1}$&$q_{1}=9.0639\times10^{-2}~\textmd{MHz}^{-1}$\\
$q_{2}=-0.16104~\textmd{MHz}^{-2}$&$q_{2}=5.1437\times10^{-2}~\textmd{MHz}^{-2}$\\
$q_{3}=2.5016\times10^{-2}~\textmd{MHz}^{-3}$&$q_{3}=-1.1041\times10^{-2}~\textmd{MHz}^{-3}$\\
$q_{4}=-1.8631\times10^{-3}~\textmd{MHz}^{-4}$&$q_{4}=8.3451\times10^{-4}~\textmd{MHz}^{-4}$\\
$q_{5}=5.3498\times10^{-5}~\textmd{MHz}^{-5}$&$q_{5}=-2.2295\times10^{-5}~\textmd{MHz}^{-5}$\\
\hline\hline
\end{tabular}
\end{table}

\section{Discussion}\label{dis}
Considering a realistic experimental setup, there are some problems that should be addressed by further discussions, such as the phase induced by wave vectors, the atomic position fluctuation, and the effectiveness of vdW interaction.
\subsection{The phases induced by wave vectors}\label{5a}

\begin{figure}
\centering\scalebox{0.32}{\includegraphics{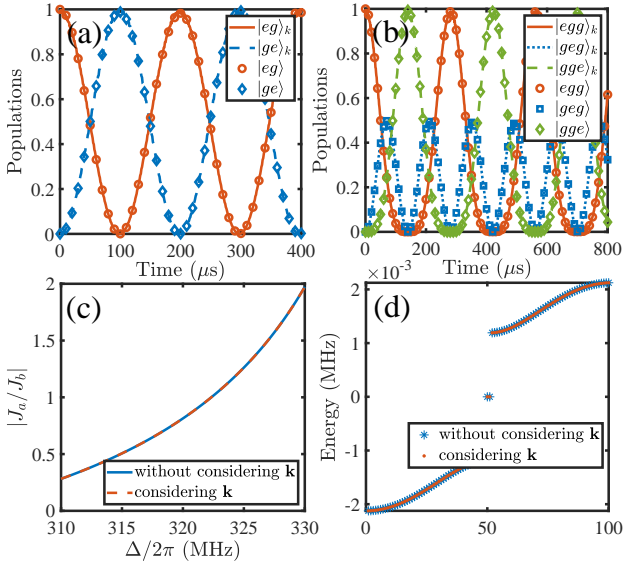}}
\caption{\label{wave}The system dynamics with and without considering the wave vectors (marked with subscript $k$ after the ket symbol). (a) and (b) correspond to the dynamics of an isometric chain structure with $N=2$ and $N=3$, respectively, where $r=4.1~\mu$m. The evolutions are governed by Eqs.~(\ref{H1}) and (\ref{WH1}). (c) The ratio of $J_{a}$ to $J_{b}$ is shown as a function of the detuning $\Delta$. (d) The energy spectrum of the SSH model with $N=100$, where $r_{2i-2,2i}=4~\mu$m and $r_{2i,2i+1}=4.1~\mu$m. The other parameters are taken as $\delta=\Omega_{p}=2\pi\times1~$MHz, $\Omega=0.05\Omega_{p}$, and $\Delta=300\Omega_{p}$.}
\end{figure}
\begin{figure}
\centering
\includegraphics[scale=0.22]{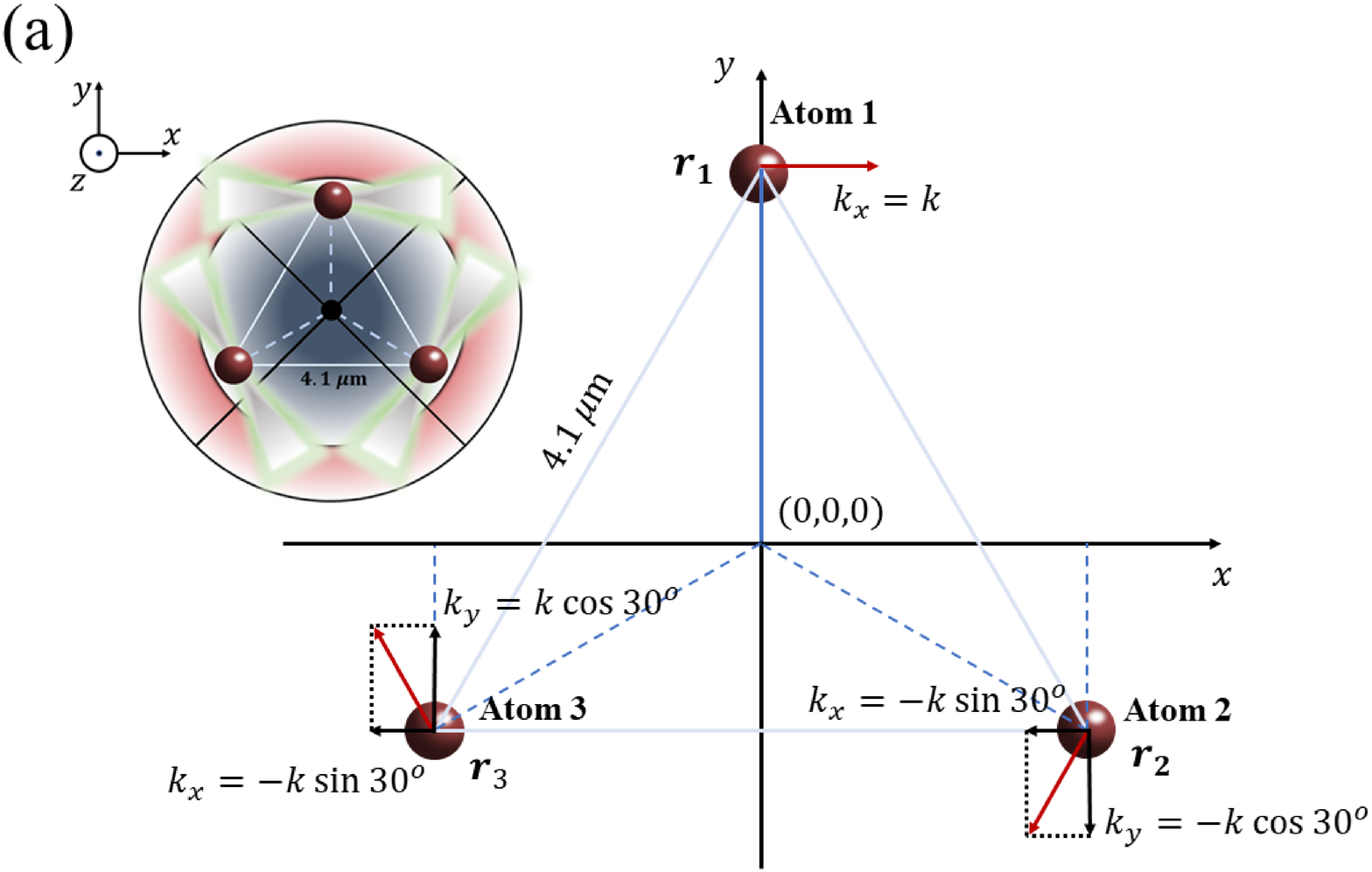}
\hspace{1in}
\includegraphics[scale=0.22]{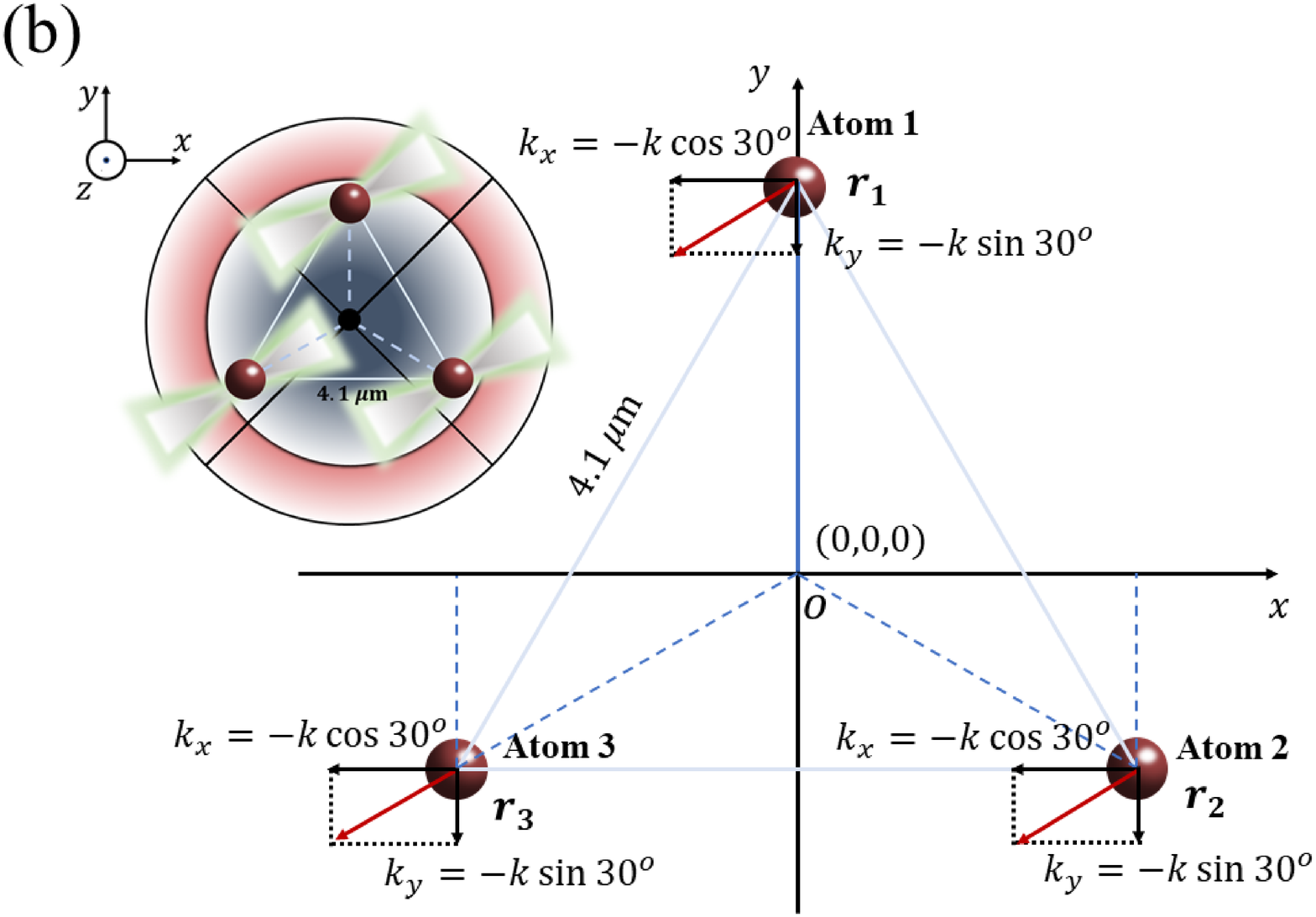}
\caption{\label{direction}The feasible experimental structure for the equilateral triangle structure, where the global laser fields $\Omega_{p1}$ and $\Omega_{p2}$ are counter-propagating along $z$-axis. (a) The effective wave vectors $\mathbf{k}_{i}$ of weak fields $\Omega_{i}$ are orthogonal to the atomic position $\mathbf{r}_{i}$. (b) The weak fields $\Omega_{i}$ propagate along a straight line perpendicular to atoms $1$ and atom $2$.}
\end{figure}

To be more intuitive, the phases caused by the wave-vectors have been ignored in above analysis. However, once the wave vectors are specified, the corresponding phases should be taken into account.

For the method of quantum state transport, $^{87}$Rb atoms are arranged in a line along the quantization $z$-axis. The phases induced by the individual beams $\Omega_{i1}$ and $\Omega_{i2}$ can be neglected, but the global laser fields $\Omega_{pi}$ propagating along $z$-axis will induce phase factors $e^{i\mathbf{k}_{pi}\cdot\mathbf{r}_{i}}$. Since the global laser beams are counter-propagating along $z$-axis, the effective wave vector introduced by $\Omega_{p}$ is $k_{z}/2\pi=|\lambda_{p1}^{-1}-\lambda_{p2}^{-1}|\simeq5.062\times10^{6}/$m.
Taking $N=2$ as an example, after denoting the position of the two atoms as $\mathbf{r}_{1}=(0, 0, z_{1})$ and $\mathbf{r}_{2}=(0, 0, z_{2})$, the system Hamiltonian of Eq.~(\ref{H1}) is rewritten as
\begin{eqnarray}\label{WH1}
H_{I}^{(N)}&=&\sum_{j=1}^{N}\Omega_j|r_j\rangle\langle g_j|+\Omega_{p}e^{ik_{z}z_{j}}|r_j\rangle\langle e_j|+{\rm H.c.}+\delta|g_{j}\rangle\langle g_j|\nonumber\\&&+\Delta|e_{j}\rangle\langle e_j|+\sum_{j<k}{\cal U}_{j k}|r_jr_k\rangle\langle r_jr_k|.
\end{eqnarray}
Meanwhile, the Hamiltonian shown in Eq.~(\ref{H21}) and Eq.~(\ref{HV}) should be modified as
\begin{eqnarray}
H_{ph}&=&\Omega_{p}(e^{ik_{z}z_{1}}|er\rangle\langle rr|+e^{ik_{z}z_{2}}|re\rangle\langle rr|)+{\rm H.c.}\nonumber\\&&-\delta(|er\rangle\langle er|+|re\rangle\langle re|+|rr\rangle\langle rr|),
\end{eqnarray}
and
\begin{equation}
V_{+}=V^{\dag}_{-}=\Omega_{1}|re\rangle\langle ge|+\Omega_{2}|er\rangle\langle eg|.
\end{equation}
Applying the effective operator method
\begin{equation}
H_{\textmd{eff}}=-\frac{1}{2}[V_{-}H_{ph}^{-1}V_{+}+V_{-}(H_{ph}^{-1})^{\dag}V_{+}],
\end{equation}
we have
\begin{equation}\label{Hwave1}
H_{\textmd{eff}}=J_{12}e^{ik_{z}(z_{2}-z_{1})}|ge\rangle\langle eg|+{\rm H.c.},
\end{equation}
where $J_{12}=\Omega_{1}\Omega_{2}\Omega_{p}^{2}/(\delta^{3}-2\delta\Omega_{p}^{2})$.
Starting from the initial state $|eg\rangle$, governed by the effective Hamiltonian of Eq.~(\ref{Hwave1}), we have $|\Psi(t)\rangle=\cos(J_{12}t)|eg\rangle-ie^{ik_{z}(z_{2}-z_{1})}\sin(J_{12}t)|ge\rangle$. The wave vector introduce an extra relative phase $e^{ik_{z}(z_{2}-z_{1})}$ between state $|eg\rangle$ and $|ge\rangle$. It is apparent from the above form that the effective phase factor caused by wave vectors is only related to the relative position of the adjacent atoms.
Therefore it is easy to obtain the effective Hamiltonian for arbitrary $N$ particles read as
\begin{equation}\label{Hwave2}
H_{\textmd{eff}}^{(N)}=\sum_{j=1}^{N-1}J_{j,j+1}e^{-ik_{z}r}\sigma_j^{+}\sigma_{j+1}^{-}+{\rm H.c.},
\end{equation}
where $J_{j,j+1}=\Omega_j\Omega_{j+1}\Omega_p^2/(\delta^{3}-2\delta\Omega_p^2)$.
Starting from the initial state $\sigma_{1}^{+}|gg...g\rangle_{N}$, extra phases $e^{i(j-1)k_{z}r}$ are introduced on states $\sigma_{j}^{+}|gg...g\rangle_{N}$. However, these relative phases do not affect the transmission of the single-excited state, as shown in Fig.~\ref{wave}(a) and \ref{wave}(b), where the system dynamics is simulated by Hamiltonian of Eq.~(\ref{WH1}) for the case of $N=2$ and 3.

From another perspective, by absorbing the phase factor $e^{-ik_{z}(z_{j+1}-z_{j})}$ into the redefined space-dependent state $|\tilde{e}_{j}\rangle=e^{ik_{z}z_{j}}|e_{j}\rangle$, the effective Hamiltonian under the new basis vectors can be written as
\begin{equation}\label{Hwave2}
H_{\textmd{eff}}^{(N)}=\sum_{j=1}^{N-1}J_{j,j+1}\tilde{\sigma}_j^{+}\tilde{\sigma}_{j+1}^{-}+{\rm H.c.},
\end{equation}
where $\tilde{\sigma}_{j}^{+}=|\tilde{e}_{j}\rangle\langle g_{j}|$, which remains to a Heisenberg $XX$ spin chain restricted in the single-excitation manifold.

For the topological model, there is a relative phase existed between $\tilde{J}_{a}$ and $\tilde{J}_{b}$ owing to various spaces between atoms which are shown as
\begin{eqnarray}
\tilde{J}_{a}&=&J_{a}e^{ikr_{2i-1,2i}},~~\tilde{J}_{b}=J_{b}e^{ikr_{2i,2i+1}}.
\end{eqnarray}
As shown in Fig.~\ref{wave}(c) and \ref{wave}(d), under the same parameters, the values of $|J_{a}/J_{b}|$ and the energy band structures of the system do not change, regardless of whether the wave vector $\mathbf{k}$ is considered or not. It's just that the forms of two edge states become $|\Psi\rangle_{\textmd{edge}}\simeq(e^{ik_{z}z_{1}}\sigma_{1}^{+}|gg...g\rangle_{N}\pm e^{ik_{z}z_{N}}\sigma_{N}^{+}|gg...g\rangle_{N})/\sqrt{2}$ for large $N$.

\begin{figure}\scalebox{0.32}{\includegraphics{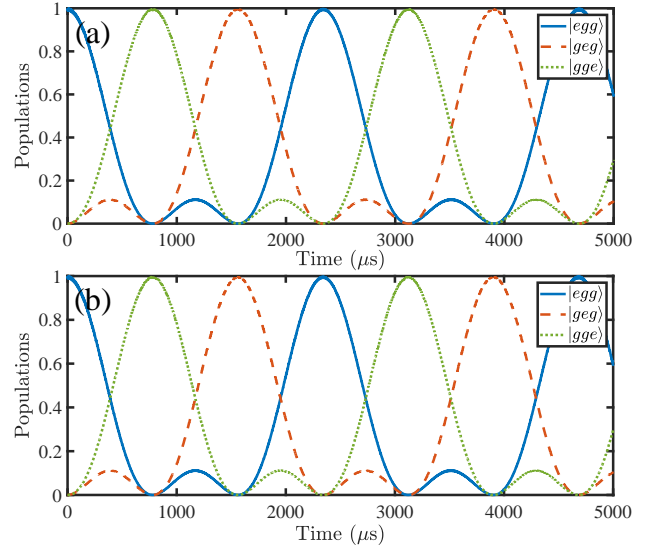}}
\caption{\label{wave_liu}The dynamics of the chiral motion of atomic excitation with $\tilde{\tau}_{ac}=0.12425~\mu$s. (a) corresponds to the experimental structure shown in Fig.~\ref{direction}(a). (b) corresponds to the experimental structure shown in Fig.~\ref{direction}(b). The other parameters are taken as $\delta=\Omega_{p}=2\pi\times1~$MHz, $\Omega=0.05\Omega_{p}$, $\Delta=300\Omega_{p}$ and $r=4.1~\mu$m.}
\end{figure}
\begin{table*}
\centering
\caption{\label{wave_phase}The effective coupling strengths and the effective phases correspond to different lighting modes with $\tilde{\tau}_{ac}=0.12425~\mu$s.}
\setlength{\tabcolsep}{2mm}
\begin{tabular}{cccccccc}
\hline\hline
The lighting mode&$J_{\textmd{eff}}^{12}~$(kHz)&$J_{\textmd{eff}}^{23}~$(kHz)&$J_{\textmd{eff}}^{31}~$(kHz)&$\Phi_{12}~(\pi)$&$\Phi_{23}~(\pi)$&$\Phi_{31}~(\pi)$&$\Phi_{z}~(\pi)$\\
\hline
Without considering $\mathbf{k}$&1.5481&1.5481&1.5479&0.1651&0.1626&0.1651&0.4928\\
Orthogonal to position $\mathbf{r}_{i}$  &1.5481&1.5481&1.5479&0.1651&0.1626&0.1651&0.4928\\
Perpendicular to atom $1$ and atom $2$&1.5481&1.5481&1.5479&0.1651&0.1133&0.2144&0.4928\\
\hline\hline
\end{tabular}
\end{table*}
\begin{figure*}\scalebox{0.35}{\includegraphics{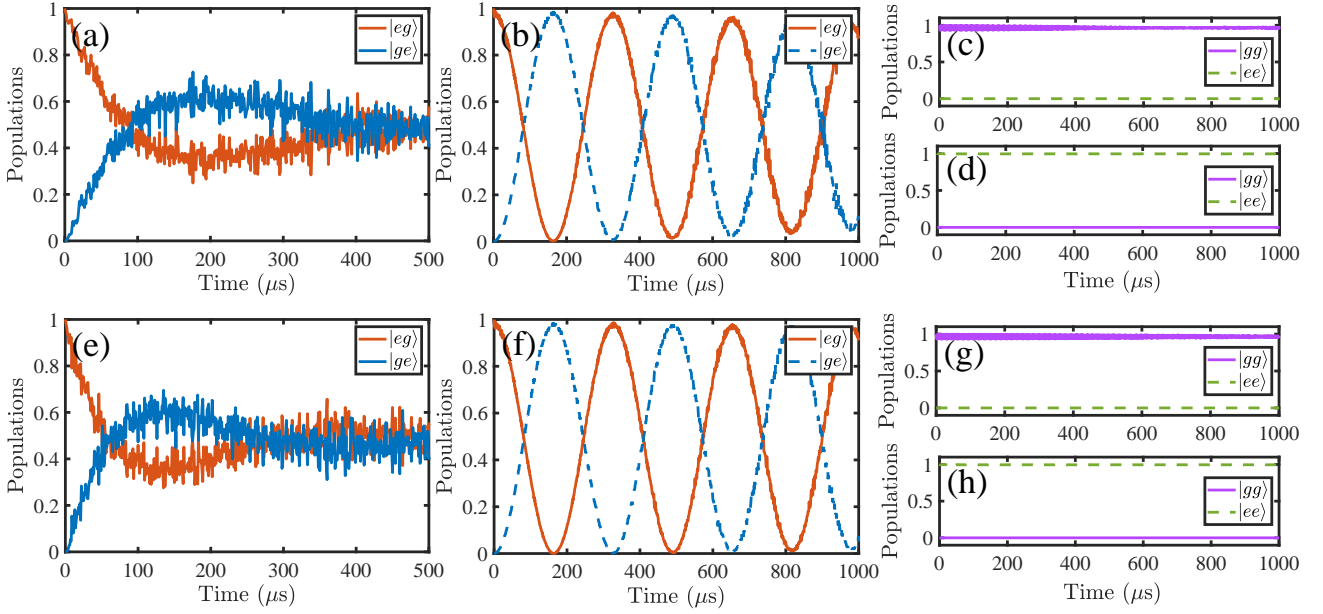}}
\caption{\label{Dr1}Transport dynamics considering the position fluctuations. (a) and (e) correspond to populations of states $|eg\rangle$ and $|ge\rangle$ together with considering the random fluctuation obeys the uniform density and standard normal density distribution, respectively. The parameters are taken as $\delta=\Omega_{p}=2\pi\times1~$MHz, $\Omega=0.05\Omega_{p}$, $\Delta=300\Omega_{p}$ and ${\cal U}'_{j,j+1}\simeq300\Omega_p+F(t)$. (b)-(d) Populations of states $|eg\rangle$, $|ge\rangle$, $|ee\rangle$ and $|gg\rangle$ under the adjusted parameters with $F(t)$ uniformly distributed, respectively. (f)-(h) The dynamics of the system with $F(t)$ obeying standard normal density distribution. The optimized parameters are taken as $\delta=\Omega_{p}=2\pi\times10~$MHz, $\Omega=0.1\Omega_{p}$, $\Delta=30\Omega_{p}$ and ${\cal U}'_{j,j+1}\simeq10\Omega_p+F(t)$.}
\end{figure*}

For the chiral motion of atomic excitation, in order to avoid introducing more relative phases, the propagation directions of the laser fields are redetermined. By adjusting the direction of the external magnetic field, the quantization $z$-axis is redefined as the direction perpendicular to the regular triangle plane. With $r=4.1~\mu$m, the radius of a circle surrounded by three atoms is about $2.4~\mu$m. The collective or independent addressing of atoms can be realized by adjusting the size of laser beam waist. Accordingly, the global laser fields $\Omega_{pi}$ propagate along $z$-axis and the local laser fields $\Omega_{i1}$ and $\Omega_{i2}$ propagate perpendicular to the $z$-axis. After defining the center position coordinates of the regular triangular plane as $(0, 0, 0)$, the phase factors are brought by the weak laser fields, while that brought by the strong fields can be ignored. As shown in Fig.~\ref{direction}(a) and \ref{direction}(b), we have discussed two situations. One of which is that the effective wave vectors $\mathbf{k}_{i}$ of weak fields $\Omega_{i}$ are orthogonal to the atomic position $\mathbf{r}_{i}$. While the other one is the weak fields $\Omega_{i}$ propagate in the same direction, such as along a straight line perpendicular to atom $1$ and atom $2$. The piecewise constant Hamiltonian of Eq.~(\ref{C1}) can be rewritten as
\begin{eqnarray}\label{CW}
H_{i}&=&\Omega_{i}e^{-i\mathbf{k}_{i}\cdot\mathbf{r}_{i}}|r_{i}\rangle\langle g_i|+\sum_{j=1}^{4}\Omega_{p}|r_{j}\rangle\langle e_j|+{\rm H.c.}
\nonumber\\&&+\delta|g_{j}\rangle\langle g_j|+\Delta|e_{j}\rangle\langle e_j|+\sum_{j<k}{\cal U}_{j k}|r_jr_k\rangle\langle r_jr_k|,
\end{eqnarray}
with effective wave vector $k_{i}\simeq5.35\times10^{6}/$m introduced by $\Omega_{i}$. Taking $\tilde{\tau}_{ac}=0.12425~\mu$s as an example, Fig.~\ref{wave_liu}(a) and \ref{wave_liu}(b) show the excited population transport between atoms corresponding to the experimental structure shown in Fig.~\ref{direction}(a) and \ref{direction}(b), respectively.
It proves that the relative phases do not affect the chiral motion of atomic excitation.
As shown in Tabel.~\ref{wave_phase}, we further research on the effective coupling strengths $J_{\textmd{eff}}^{12(23,31)}$ and the effective phases $\Phi_{12(23,31)}$ in the above two cases.
When the wave vectors $\mathbf{k}_{i}$ of $\Omega_{i}$ are orthogonal to the atomic position $\mathbf{r}_{i}$, we have $\mathbf{k}_{i}\cdot\mathbf{r}_{i}=0$. It equals to the method without considering $\mathbf{k}$.
In contrast, for the case that the weak fields $\Omega_{i}$ propagate in the same direction, such as along a straight line perpendicular to atom $1$ and atom $2$, the effective phases will change. However, the total phase $\Phi_{z}$ of the system is basically unchanged, which ensures the chiral motion of atomic excitation.
\begin{figure}\scalebox{0.32}{\includegraphics{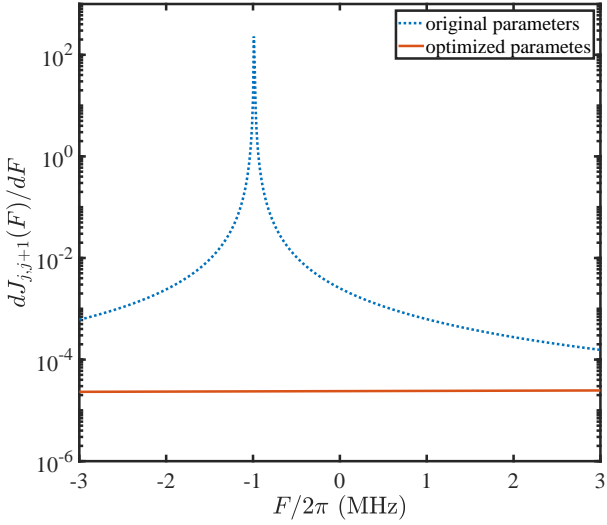}}
\caption{\label{DU_Jeff}The derivation of $J_{j,j+1}(F)$ to $F$ under the original and optimized parameters. The original parameters are taken as $\delta=\Omega_{p}=2\pi\times1~$MHz, $\Omega=0.05\Omega_{p}$, $\Delta=300\Omega_{p}$ and ${\cal U}_{j,j+1}=300\Omega_p$. The optimized parameters are taken as $\delta=\Omega_{p}=2\pi\times10~$MHz, $\Omega=0.1\Omega_{p}$, $\Delta=30\Omega_{p}$, and ${\cal U}_{j,j+1}=10\Omega_p$.}
\end{figure}

\subsection{The atomic position fluctuation}\label{5b}

In Sec.~\ref{II}, we only discuss the impact of systematic error with a  fixed atomic position. In fact, the atoms have a spacial extent, which gives rise to fluctuation in position and results in a random fluctuation on the vdW interaction. We repeat numerical simulation of this stochastic process 50 times and average the results which are shown in Fig.~\ref{Dr1}.

Considering the random fluctuations, the vdW interaction between nearest neighbour atoms can be rewritten as ${\cal U}'_{j,j+1}(t)={\cal U}_{j,j+1}+F(t)$, where $F(t)$ is assumed as uniform distributed on the interval $[-a,a]$, which is decided by the fluctuation $\delta r$ of relative distance caused by random motion of atoms, shown as $F(t)/2\pi=-a+2a\xi(t)~$MHz. In which, $\xi(t)$ is an uniformly distributed random numbers in the interval $[0,1]$.
With $F(t)$ changing every microsecond, the quantum state transport is damaged under the original parameters $\delta=\Omega_p=2\pi\times1~$MHz, $\Omega=0.05\Omega_p$ and ${\cal U}_{j,j+1}=\Delta=300\Omega_p$ with $a=3$ as shown in Fig.~\ref{Dr1}(a). The corresponding fluctuation of relative distance is about $\delta r\simeq0.01\mu$m.
In order to solve this problem and weaken the influence induced by fluctuation, we take out the condition ${\cal U}_{j,j+1}=\Delta$ and adjust the parameters as $\delta=\Omega_p=2\pi\times10~$MHz, $\Omega=0.1\Omega_p$, $\Delta=30\Omega_p$ and ${\cal U}_{j,j+1}=10\Omega_{p}$. The corresponding dynamics of the system after optimization is shown in Fig.~\ref{Dr1}(b).
The distinct Rabi oscillation of populations between states $|eg\rangle$ and $|ge\rangle$ can be observed.
In addition, we perform another simulation with $F(t)$ obeying the standard normal distribution shown as $F(t)/2\pi=\sqrt{-2\textmd{ln}\xi_{1}(t)}\textmd{cos}[2\pi\xi_{2}(t)]~$MHz with $\delta r\simeq0.01\mu$m.
The dynamics under original and optimized parameters are shown in Fig.~\ref{Dr1}(e) and \ref{Dr1}(f), respectively. Besides, the evolutions of states $|ee\rangle$ and $|gg\rangle$ under the optimized parameters with $F(t)$ considered as uniform and standard normal distribution are respectively shown in Fig.~\ref{Dr1}(c)[(d)] and \ref{Dr1}(g)[(h)], which prove that the dynamics of the system accord with the Heisenberg $XX$ spin chain for two particles.

The reason why the scheme based on the optimized parameters is more robust against the position fluctuation can be clearly understood by analyzing the relationship between the effective coupling strength and the change of atomic position. The effective coupling strength holds the same form as Eq.~(\ref{du1}) with ${\cal U}_{j,j+1}(F)={\cal U}_{j,j+1}+F$ where $F/2\pi\in[-3,3]~$MHz represents the variation of the vdW interaction caused by the change of atomic position.
Thus we have the derivation of $J_{j,j+1}(F)$ to $F$
\begin{equation}
\frac{dJ_{j,j+1}(F)}{dF}=\frac{(\delta-\Delta)^2\Omega^2\Omega_{p}^2}{[\delta(\delta^2-\eta_{j,j+1})-(\delta^2-2\Omega_{p}^2){\cal U}_{j,j+1}(F)]^2},
\end{equation}
where $\eta_{j,j+1}=4\Omega_{p}^{2}+\Delta^{2}-{\cal U}_{j,j+1}(F)\Delta$. As shown in Fig.~\ref{DU_Jeff}, $J_{j,j+1}(F)$ changes more dramatically with $F$ under the original parameters (there is a singularity which can be found directly from Eq.~(\ref{ED4}) as ${\cal U}_{j,j+1}=-\delta$). By contrast, under the optimized parameters, ${dJ_{j,j+1}(F)}/{dF}$ is
close to zero, which guarantees that the system dynamics is more robust against the fluctuation of the vdW interaction.
\begin{figure}
\centering\scalebox{0.32}{\includegraphics{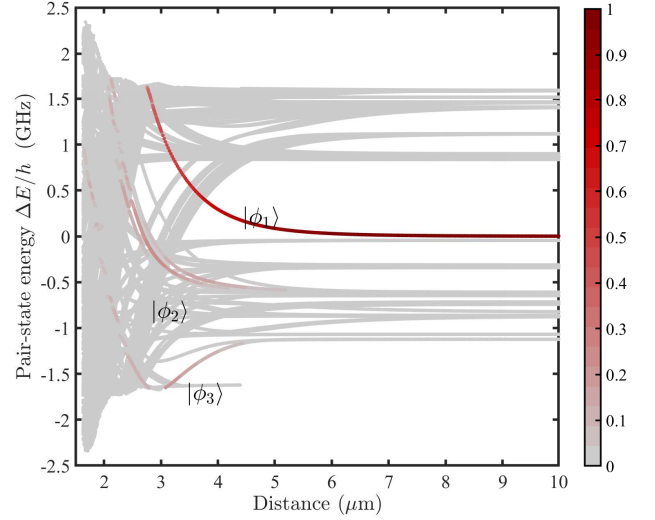}}
\caption{\label{vd}The pair potentials of $^{87}\textmd{Rb}$ atoms around the defined pair state $|rr\rangle=|73S_{1/2},m_{j}=1/2;73S_{1/2},m_{j}=1/2\rangle$. The red color denotes the overlap of the eigenstates with $|rr\rangle$.}
\end{figure}

\subsection{The effectiveness of vdW interaction}
Strictly speaking, the perturbative calculation-based estimation of the short-range vdW interaction intensity between Rydberg states is not working at all, because splittings between energy levels are smaller than interaction energies. In order to find a more practical system parameter, we rewritten the Hamiltonian of the system as
\begin{eqnarray}
H_{I}^{(N)}&=&\sum_{j=1}^{N}\Omega_j|r_j\rangle\langle g_j|+\Omega_{p}|r_j\rangle\langle e_j|+{\rm H.c.}+\delta|g_{j}\rangle\langle g_j|\nonumber\\&&+\Delta|e_{j}\rangle\langle e_j|+H_{int}^{rr},
\end{eqnarray}
where $H_{int}^{rr}$ presents the interactions between Rydberg state $|r\rangle=|73S_{1/2}\rangle$ and the states with similar energy and quantum numbers.
After diagonalizing the Hamiltonian of these quantum states  (principal quantum number $|n-73|\leq5$ and azimuthal quantum number $|L|\leq5$) based on the open software ``Alkali-Rydberg-Calculator" \cite{SIBALIC2017319}, the energy map is shown in Fig.~\ref{vd} and the population of the state $|rr\rangle=|73S_{1/2}$, $m_{j}=1/2$; $73S_{1/2}$, $m_{j}=1/2\rangle$ in the diagonalized state increases as the red color deepens.
We focus on the three eigenstates denoted as $|\phi_{1}\rangle$, $|\phi_{2}\rangle$ and $|\phi_{3}\rangle$ in Fig.~\ref{vd} with highest population of $|rr\rangle$, where $|\phi_{1}\rangle$ is mainly constructed by $\{|rr\rangle$, $|72P_{3/2}$,$m_{j}=1/2$; $73P_{3/2}$, $m_{j}=1/2\rangle$, $|73P_{3/2}$, $m_{j}=1/2$; $72P_{3/2}$, $m_{j}=1/2\rangle\}$, $|\phi_{2}\rangle$ is mainly constructed by $\{|rr\rangle$, $|73P_{3/2}$, $m_{j}=1/2$; $72P_{3/2}$, $m_{j}=1/2\rangle$, $|72P_{3/2}$,$m_{j}=1/2$; $73P_{3/2}$, $m_{j}=1/2\rangle$, $|74S_{1/2}$, $m_{j}=1/2$; $72S_{1/2}$, $m_{j}=1/2\rangle$, $|72S_{1/2}$, $m_{j}=1/2$; $74S_{1/2}$, $m_{j}=1/2\rangle\}$, while $|\phi_{3}\rangle$ is mainly constructed by $\{|rr\rangle$, $|74S_{1/2}$, $m_{j}=1/2$; $72S_{1/2}$, $m_{j}=1/2\rangle$, $|72S_{1/2}$,$m_{j}=1/2$; $74S_{1/2}$, $m_{j}=1/2\rangle$, $|73P_{1/2}$, $m_{j}=1/2$; $72P_{1/2}$, $m_{j}=1/2\rangle\}$.
Then we have $H_{int}^{rr}\approx E_{1}|\phi_{1}\rangle\langle\phi_{1}|+E_{2}|\phi_{2}\rangle\langle\phi_{2}|+E_{3}|\phi_{3}\rangle\langle\phi_{3}|$ and $|rr\rangle\approx\alpha_{1}|\phi_{1}\rangle+\alpha_{2}|\phi_{2}\rangle+\alpha_{3}|\phi_{3}\rangle$, where $\alpha_{1,2,3}$ are the probability amplitudes of state $|rr\rangle$.
Taking into account the large detuning and weak coupling strength $\alpha_{2(3)}\Omega_{p}$ between $\{|er\rangle,|re\rangle\}$ and $\{|\phi_{2}\rangle,|\phi_{3}\rangle\}$, only $|\phi_{1}\rangle$ makes contribution and $\{|\phi_{2}\rangle,|\phi_{3}\rangle\}$ can be neglected as high-frequency terms.

\begin{figure}\scalebox{0.32}{\includegraphics{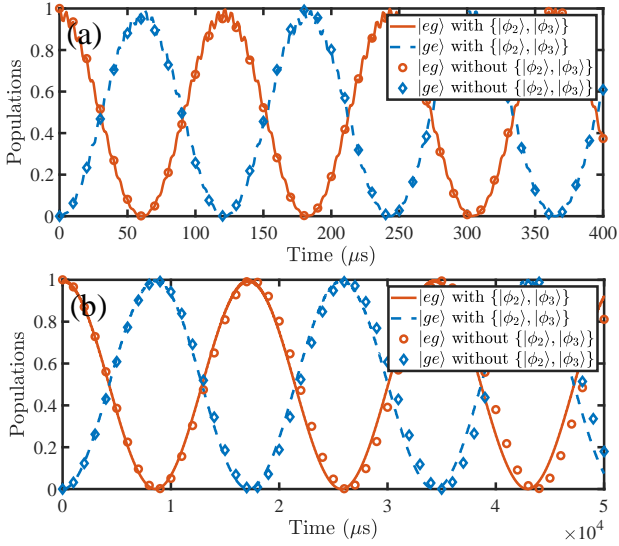}}
\caption{\label{vdW}Transport dynamics of states $|eg\rangle$ (red) and $|ge\rangle$ (blue) with and without states $|\phi_{2}\rangle$, $|\phi_{3}\rangle$. (a) corresponds to the system with $\Delta U=0$. (b) corresponds to the system with $\Delta U=-2\pi\times50~$MHz. The other parameters are taken as $\Omega_{j}=0.05\Omega_{p}$, $E_{1}=2\pi\times300~$MHz, $E_{2}=-2\pi\times511.25~$MHz, $E_{3}=-2\pi\times1258.83~$MHz, $\alpha_{1}=\sqrt{0.72}$, $\alpha_{2}=\sqrt{0.126}$, and $\alpha_{3}=\sqrt{0.088}$.}
\end{figure}

The dynamics of systems with and without considering $\{|\phi_{2}\rangle, |\phi_3\rangle\}$ are shown in Fig.~\ref{vdW} for $r_{j,j+1}=3.99~\mu$m. Fig.~\ref{vdW}(a) corresponds to the system satisfying the resonance condition $E_{1}=\Delta$, while \ref{vdW}(b) includes the deviation of unconventional Rydberg pumping conditions with $E_{1}\neq\Delta$ and $\Delta U=E_{1}-\Delta=-2\pi\times50~$MHz.
The other parameters are taken as $\delta=\Omega_{p}=2\pi\times1~$MHz, $\Omega_{j}=0.05\Omega_{p}~$MHz, $E_{1}=2\pi\times300~$MHz, $E_{2}=-2\pi\times511.25~$MHz, $E_{3}=-2\pi\times1258.83~$MHz, $\alpha_{1}=\sqrt{0.72}$, $\alpha_{2}=\sqrt{0.126}$, and $\alpha_{3}=\sqrt{0.088}$.
The corresponding numerical results prove that within a certain range of detuning $\Delta U$, $\{|\phi_{2}\rangle, |\phi_3\rangle\}$ can always be safely neglected.
Incorporating the effective operator method with the above analysis, we obtain the effective coupling strength with and without $\Delta U$ as
\begin{equation}\label{CE1}
J_{j,j+1}=\frac{\Omega_j\Omega_{j+1}\alpha_{1}^{2}\Omega_p^2}{\delta^{3}-2\delta\alpha_{1}^{2}\Omega_p^2},
\end{equation}
and
\begin{equation}\label{Jeffxiu}
J_{j,j+1}=\frac{\Omega_j\Omega_{j+1}\Omega_p^2\zeta}{\delta^{2}(\delta-\Delta U)(\delta-\Delta)-2\zeta\delta\Omega_{p}^{2}},
\end{equation}
where $\zeta=\delta+\alpha_{1}^{2}\delta-\alpha_{1}^{2}\Delta-\Delta U$.
The correction shown in Eq.~(\ref{CE1}) and (\ref{Jeffxiu}) will not affect the conclusions we obtained before.
For perfect quantum state transfer protocol and the chiral motion of atomic excitation, by comparing the forms of $J_{j, j+1}$ in Eqs.~(\ref{CE1}) and (\ref{Ar1}), it can be found that the system dynamics keeps unchanged by setting $\Omega_{p}=2\pi\times(1/\alpha_{1})~$MHz.
While for the topological model, the corresponding topological phase can always be achieved by modulating the parameter $\Delta$ and the atomic separations.
\section{Conclusion}\label{VI}
In conclusion, we have proposed a theoretical framework for studying quantum state transfer scheme inside ground-state manifold of neutral atoms by only combining  the diagonal vdW interaction with unconventional Rydberg pumping condition. The scheme successfully realized the Heisenberg $XX$ spin-chain dynamics restricted in the single-excitation manifold. Meanwhile, depending on the choice of parameters, the system dynamics can be equivalent to a second-order process only related to weak fields and the perfect quantum state transfer is realized by simply regulating the weak fields of atoms. A 1D SSH model is then constructed by differing the distance between atoms, and the system can be flexibly changed from topological trivial phase to nontrivial phase by adjusting the detuning $\Delta$. Finally, a method to realize the chiral motion of atomic excitation is provided in the equilateral triangle structure. A total flux $\Phi_{z}=\pm\pi/2$ can be obtained via periodically modulating the weak pulses without introducing any other external fields. In a word, we can get abundant physical pictures by using such a simple physical system, and we hope that our work may pave a new avenue for quantum simulation of neutral atomic system.

\section*{Acknowledgement}
The author would like to thank Dr. Kuan Zhang and Prof. Lin Li for helpful comments and suggestions. The anonymous reviewers are also
thanked for constructive comments that helped in improving
the quality of this paper.
This work is supported by National Natural Science Foundation of China (NSFC) under Grants No. 11774047 and No. 12174048. J.B.Y. acknowledges the supports from the National Research Foundation Singapore (QEP-SF1) and A*STAR Career Development Award (SC23/21-8007EP). W.L. acknowledges
support from the EPSRC through Grant No. EP/R04340X/1
via the QuantERA project ``ERyQSenS," the Royal Society
Grant No. IEC$\verb|\|$NSFC$\verb|\|$181078.

\appendix
\section{The $\sqrt{\textmd{SWAP}}$ gate}
\begin{figure}
\centering\scalebox{0.32}{\includegraphics{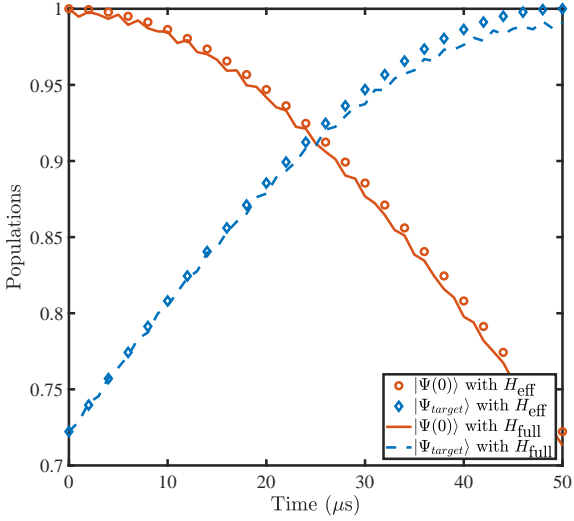}}
\caption{\label{swap}Transport dynamics of diatomic model governed by the full Hamiltonian of Eq.~(\ref{Hsw1}) after shifting the energy of $|e_{i}\rangle$ and $|g_{i}\rangle$ and the effective Hamiltonian of Eq.~(\ref{swe}), respectively. The initial state is $|\Psi(0)\rangle=1/\sqrt{3}(|eg\rangle+|gg\rangle+|ee\rangle)$ and the parameters are taken as $\delta=\Omega_{p}=2\pi\times1~$MHz, $\Omega=0.05\Omega_{p}$, and $\Delta=300\Omega_{p}$.}
\end{figure}

For a diatomic model, our method can also be used to realize the $\sqrt{\textmd{SWAP}}$ gate shown as
\begin{equation}\
\sqrt{\textmd{SWAP}}=\left[
\begin{array}{cccc}
1 & 0 & 0 & 0\\
0 & \frac{1}{2}(1+i) & \frac{1}{2}(1-i) & 0\\
0 & \frac{1}{2}(1-i) & \frac{1}{2}(1+i) & 0\\
0 & 0 & 0 & 1\\
\end{array}
\right].
\end{equation}
In the interaction picture, the Hamiltonian of the system reads
\begin{eqnarray}\label{Hsw1}
H_{I}^{(2)}&=&\sum_{j=1}^{2}\Omega_j e^{i\delta t}|r_j\rangle\langle g_j|+\Omega_{p}e^{i\Delta t}|r_j\rangle\langle e_j|+{\rm H.c.}\nonumber\\&&+{\cal U}|rr\rangle\langle rr|,
\end{eqnarray}
and the effective Hamiltonian is
\begin{eqnarray}\label{sw1}
H_{\textmd{eff}}&=&J_{\textmd{eff}}|ge\rangle\langle eg|+{\rm H.c.}+S_{1}(|ge\rangle\langle ge|+|eg\rangle\langle eg|)\nonumber\\&&+S_{2}|gg\rangle\langle gg|+S_{3}|ee\rangle\langle ee|,
\end{eqnarray}
where $J_{\textmd{eff}}=\Omega_{1}\Omega_{2}\Omega_{p}^{2}/(\delta^{3}-2\delta\Omega_{p}^{2})$, $S_{1}=\Omega_{1}\Omega_{2}(\delta^{2}-\Omega_{p}^{2})/(\delta^{3}-2\delta\Omega_{p}^{2})+\Omega_{p}^{2}/\Delta$, $S_{2}=(\Omega_{1}^{2}+\Omega_{2}^{2})/\delta$, $S_{3}=2\Omega_{p}^{2}/\Delta$. The energy of states $|ee\rangle$ and $|gg\rangle$ can be shifted to zero by Stark shifts through coupling to extra states off-resonantly. Under the condition $\delta=\Omega_{p}$ and $\Omega_{1}=-\Omega_{2}=\Omega$, the effective Hamiltonian can be simplified as
\begin{equation}\label{swe}
H_{\textmd{eff}}=\frac{\Omega^{2}}{\delta}(|eg\rangle\langle ge|+|ge\rangle\langle eg|)-\frac{\Omega^{2}}{\delta}(|eg\rangle\langle eg|+|ge\rangle\langle ge|).
\end{equation}
The system governed by this Hamiltonian can realize a quantum $\sqrt{\textmd{SWAP}}$ gate with $t=\delta\pi/4\Omega^{2}$. Corresponding population evolutions under the Hamiltonian of Eq.~(\ref{Hsw1}) and Eq.~(\ref{swe}) are respectively shown in Fig.~\ref{swap}, where the parameters are taken as  $\delta=\Omega_{p}=2\pi\times1~$MHz, $\Omega=0.05\Omega_{p}$, and $\Delta=300\Omega_{p}$. At $t=50~\mu$s, the population of state $1/\sqrt{3}(|eg\rangle+|gg\rangle+|ee\rangle)$ is transferred to $|\Psi_{target}\rangle=1/\sqrt{3}[1/2(1+i)|eg\rangle+1/2(1-i)|ge\rangle+|gg\rangle+|ee\rangle]$.
\section{Building block for Chiral motion of atomic excitation}
To construct the chiral motion of atomic excitation, we introduce the piecewise Hamiltonian.
\begin{equation}\label{BEc2}
 H(t)=\left\{
              \begin{array}{lr}
              H_{1},\ \ t\in[0, T/3)\\
              H_{2},\ \ t\in[T/3, 2T/3)\\
              H_{3},\ \ t\in[2T/3, T)\\
              \end{array}
\right.
\end{equation}
where
\begin{eqnarray}\label{EE1}
H_{i}&=&\Omega_{i}e^{-i\delta t}|r_{i}\rangle\langle g_i|+\sum_{j=1}^{3}\Omega_{p}e^{-i\Delta t}|r_{j}\rangle\langle e_j|+{\rm H.c.}\nonumber\\&&+\sum_{j<k}{\cal U}_{j k}|r_jr_k\rangle\langle r_jr_k|.
\end{eqnarray}
Since $H_{i} (i=1,2,3)$ are not commuted with each other, this periodical driving will induce the effective coupling strengths associating with the period $T$ and the phases between ground states. Because the computational space is very large, we calculate the effective Hamiltonian by bonding analytic and numerical methods together. For a given $\tau$, the system Hamiltonian can be numerically obtained as a large matrix and the effective Hamiltonian can be expressed in logarithmic form
\begin{equation}\label{EfH}
H_{\textmd{eff}}=\frac{i}{T}\textmd{ln}(e^{-iH_{3}\tau}e^{-iH_{2}\tau}e^{-iH_{1}\tau}).
\end{equation}
According to Eq.~(\ref{BEc2}), under the unconventional Rydberg pumping condition and in the limit of $\Delta\gg\Omega_{p}$, the high oscillating term proportional to $\Delta$ can be neglected. Thus the dynamics is restricted in states $|egg\rangle$, $|egr\rangle$, $|rgr\rangle$, $|rge\rangle$, $|gge\rangle$, $|gre\rangle$, $|grr\rangle$, $|ger\rangle$, $|geg\rangle$, $|reg\rangle$, $|rrg\rangle$, and $|erg\rangle$. To simplify computational space, we only consider the subspace constructed by these 12 states in the subsequent calculations.

\begin{figure*}\scalebox{0.35}{\includegraphics{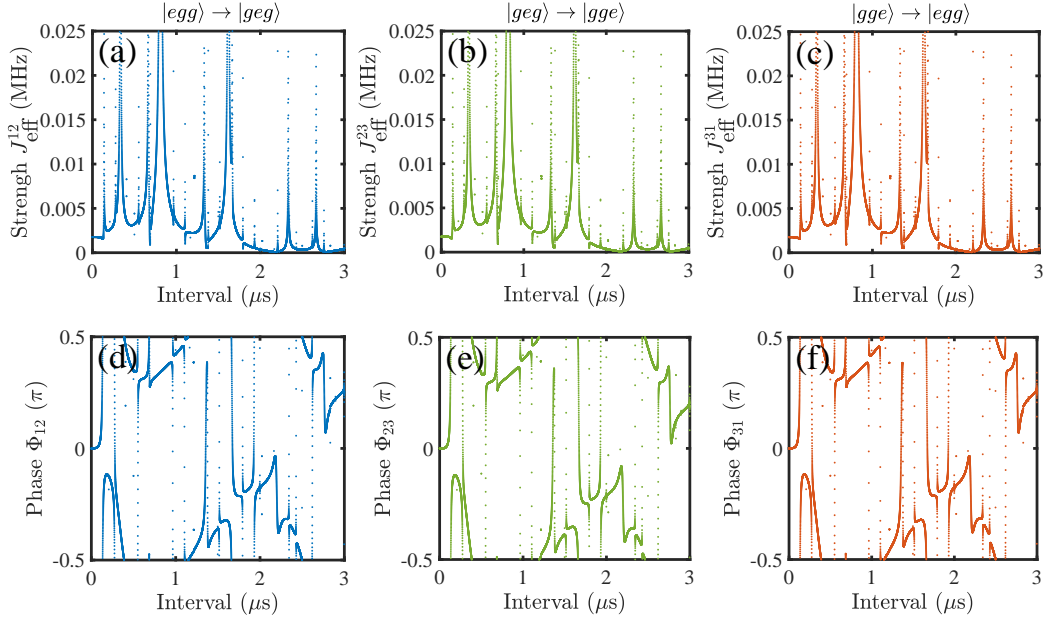}}
\caption{\label{Phase}The effective coupling strengths and phases between arbitrary ground states induced by periodical driving. The parameters are taken as $\delta=\Omega_{p}=2\pi\times1~$MHz, $\Omega_{i}=0.05\Omega_{p}$, and $\Delta=2\pi\times300~$MHz.}
\end{figure*}
In order to obtain the specific value of phases, we first numerically expend the effective Hamiltonian as a 12 $\times$ 12 matrix with some definite value of $\tau$ read as
\begin{widetext}
\begin{equation}\label{Es0}
 H=\left[
              \begin{array}{c|ccc|c|ccc|c|ccc}
              \rho_{1,1} & \rho_{1,2} & \rho_{1,3} & \rho_{1,4} & \rho_{1,5}
              & \rho_{1,6} & \rho_{1,7} & \rho_{1,8} & \rho_{1,9} & \rho_{1,10} & \rho_{1,11} & \rho_{1,12}\\
              \hline
              \rho_{1,2}^{*} & \rho_{2,2} & \rho_{2,3} & \rho_{2,4} & \rho_{2,5}
              & \rho_{2,6} & \rho_{2,7} & \rho_{2,8} & \rho_{2,9} & \rho_{2,10} & \rho_{2,11} & \rho_{2,12}\\
              \rho_{1,3}^{*} & \rho_{2,3}^{*} & \rho_{3,3} & \rho_{3,4} & \rho_{3,5}
              & \rho_{3,6} & \rho_{3,7} & \rho_{3,8} & \rho_{3,9} & \rho_{3,10} & \rho_{3,11} & \rho_{3,12}\\
              \rho_{1,4}^{*} & \rho_{2,4}^{*} & \rho_{3,4}^{*} & \rho_{4,4} & \rho_{4,5}
              & \rho_{4,6} & \rho_{4,7} & \rho_{4,8} & \rho_{4,9} & \rho_{4,10} & \rho_{4,11} & \rho_{4,12}\\
              \hline
              \rho_{1,5}^{*} & \rho_{2,5}^{*} & \rho_{3,5}^{*} & \rho_{4,5}^{*} & \rho_{5,5}
              & \rho_{5,6} & \rho_{5,7} & \rho_{5,8} & \rho_{5,9} & \rho_{5,10} & \rho_{5,11} & \rho_{5,12}\\
              \hline
               \rho_{1,6}^{*} & \rho_{2,6}^{*} & \rho_{3,6}^{*} & \rho_{4,6}^{*} & \rho_{5,6}^{*}
              & \rho_{6,6} & \rho_{6,7} & \rho_{6,8} & \rho_{6,9} & \rho_{6,10} & \rho_{6,11} & \rho_{6,12}\\
               \rho_{1,7}^{*} & \rho_{2,7}^{*} & \rho_{3,7}^{*} & \rho_{4,7}^{*} & \rho_{5,7}^{*}
              & \rho_{6,7}^{*} & \rho_{7,7} & \rho_{7,8} & \rho_{7,9} & \rho_{7,10} & \rho_{7,11} & \rho_{7,12}\\
               \rho_{1,8}^{*} & \rho_{2,8}^{*} & \rho_{3,8}^{*} & \rho_{4,8}^{*} & \rho_{5,8}^{*}
              & \rho_{6,8}^{*} & \rho_{7,8}^{*} & \rho_{8,8} & \rho_{8,9} & \rho_{8,10} & \rho_{8,11} & \rho_{8,12}\\
              \hline
               \rho_{1,9}^{*} & \rho_{2,9}^{*} & \rho_{3,9}^{*} & \rho_{4,9}^{*} & \rho_{5,9}^{*}
              & \rho_{6,9}^{*} & \rho_{7,9}^{*} & \rho_{8,9}^{*} & \rho_{9,9} & \rho_{9,10} & \rho_{9,11} & \rho_{9,12}\\
              \hline
               \rho_{1,10}^{*} & \rho_{2,10}^{*} & \rho_{3,10}^{*} & \rho_{4,10}^{*} & \rho_{5,10}^{*}
              & \rho_{6,10}^{*} & \rho_{7,10}^{*} & \rho_{8,10}^{*} & \rho_{9,10}^{*} & \rho_{10,10} & \rho_{10,11} & \rho_{10,12}\\
               \rho_{1,11}^{*} & \rho_{2,11}^{*} & \rho_{3,11}^{*} & \rho_{4,11}^{*} & \rho_{5,11}^{*}
              & \rho_{6,11}^{*} & \rho_{7,11}^{*} & \rho_{8,11}^{*} & \rho_{9,11}^{*} & \rho_{10,11}^{*} & \rho_{11,11} & \rho_{11,12}\\
               \rho_{1,12}^{*} & \rho_{2,12}^{*} & \rho_{3,12}^{*} & \rho_{4,12}^{*} & \rho_{5,12}^{*}
              & \rho_{6,12}^{*} & \rho_{7,12}^{*} & \rho_{8,12}^{*} & \rho_{9,12}^{*} & \rho_{10,12}^{*} & \rho_{11,12}^{*} & \rho_{12,12}\\
              \end{array}
\right].
\end{equation}
\end{widetext}

\begin{figure*}\scalebox{0.35}{\includegraphics{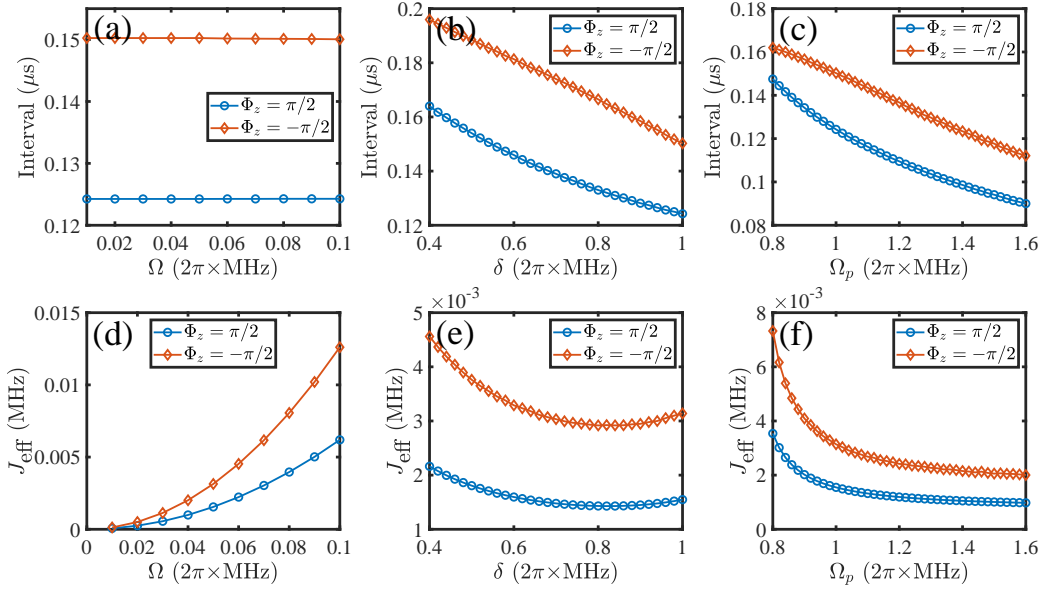}}
\caption{\label{Tau_para}(a) and (d) respectively shows the change of $\tilde{\tau}$ and $J_{\textmd{eff}}$ with $\Omega$. The other parameters are taken as $\delta=\Omega_{p}=2\pi\times1~$MHz, $\Delta=300\Omega_{p}$. (b) and (e) respectively shows the change of $\tilde{\tau}$ and $J_{\textmd{eff}}$ with $\delta$. The other parameters are taken as $\Omega_{p}=2\pi\times1~$MHz, $\Omega=2\pi\times0.05~$MHz, $\Delta=300\Omega_{p}$. (c) and (f) respectively show the change of $\tilde{\tau}$ and $J_{\textmd{eff}}$ with $\Omega_{p}$. The other parameters are taken as $\delta=2\pi\times1~$MHz, $\Omega=2\pi\times0.05~$MHz, $\Delta=2\pi\times300~$MHz.}
\end{figure*}

The order of basis vectors is taken as $|egg\rangle$, $|erg\rangle$, $|rrg\rangle$, $|reg\rangle$, $|geg\rangle$, $|ger\rangle$, $|grr\rangle$, $|gre\rangle$, $|gge\rangle$, $|rge\rangle$, $|rgr\rangle$, and $|egr\rangle$. Then the Hamiltonian corresponding to the coupling between arbitrary two ground states respectively read as
\begin{equation}\label{Es1}
 H_{\textmd{I1}}=\left[
              \begin{array}{c|ccc|c}
              \rho_{1,1} & \rho_{1,2} & \rho_{1,3} & \rho_{1,4} & \rho_{1,5}\\
              \hline
              \rho_{1,2}^{*} & \rho_{2,2} & \rho_{2,3} & \rho_{2,4} & \rho_{2,5}\\
              \rho_{1,3}^{*} & \rho_{2,3}^{*} & \rho_{3,3} & \rho_{3,4} & \rho_{3,5}\\
              \rho_{1,4}^{*} & \rho_{2,4}^{*} & \rho_{3,4}^{*} & \rho_{4,4} & \rho_{4,5}\\
              \hline
              \rho_{1,5}^{*} & \rho_{2,5}^{*} & \rho_{3,5}^{*} & \rho_{4,5}^{*} & \rho_{5,5}
              \end{array}
\right],
\end{equation}
\begin{equation}
 H_{\textmd{I2}}=\left[
              \begin{array}{c|ccc|c}
              \rho_{5,5} & \rho_{5,6} & \rho_{5,7} & \rho_{5,8} & \rho_{5,9}\\
              \hline
              \rho_{5,6}^{*} & \rho_{6,6} & \rho_{6,7} & \rho_{6,8} & \rho_{6,9}\\
              \rho_{5,7}^{*} & \rho_{6,7}^{*} & \rho_{7,7} & \rho_{7,8} & \rho_{7,9}\\
              \rho_{5,8}^{*} & \rho_{6,8}^{*} & \rho_{7,8}^{*} & \rho_{8,8} & \rho_{8,9}\\
              \hline
              \rho_{5,9}^{*} & \rho_{6,9}^{*} & \rho_{7,9}^{*} & \rho_{8,9}^{*} & \rho_{9,9}
              \end{array}
\right],
\end{equation}
\begin{equation}
 H_{\textmd{I3}}=\left[
              \begin{array}{c|ccc|c}
              \rho_{9,9} & \rho_{9,10} & \rho_{9,11} & \rho_{9,12} & \rho_{1,9}^{*}\\
              \hline
              \rho_{9,10}^{*} & \rho_{10,10} & \rho_{10,11} & \rho_{10,12} & \rho_{1,10}^{*}\\
              \rho_{9,11}^{*} & \rho_{10,11}^{*} & \rho_{11,11} & \rho_{11,12} & \rho_{1,11}^{*}\\
              \rho_{9,12}^{*} & \rho_{10,12}^{*} & \rho_{11,12}^{*} & \rho_{12,12} & \rho_{1,12}^{*}\\
              \hline
              \rho_{1,9} & \rho_{1,10} & \rho_{1,11} & \rho_{1,12} & \rho_{1,1}
              \end{array}
\right].
\end{equation}
Combined with above three 5 $\times$ 5 matrices, the effective couplings between any two ground states of $\{|egg\rangle$, $|geg\rangle$, $|gge\rangle\}$ can be obtained via second-order perturbation theory. Taking $H_{\textmd{I1}}$ as an example, the specified calculation process is illustrated below. First we diagonalized the strong coupling part which correspond to the 3 by 3 matrix in the middle of $H_{\textmd{I1}}$. Then the Hamiltonian of this part can be represented by the its eigenvalues and eigenvectors as
\begin{figure}
\centering\scalebox{0.32}{\includegraphics{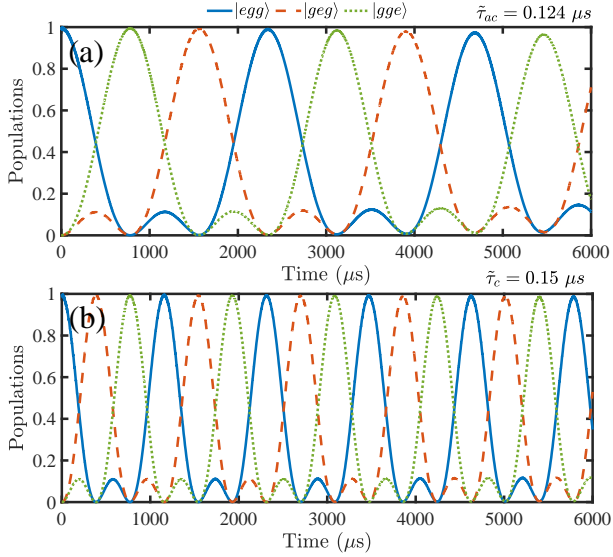}}
\caption{\label{tau}Populations of ground states as a function of time corresponding to  $\Phi_{z}=\pi/2, -\pi/2$ while the time interval is respectively taken as $0.124~\mu$s and $0.15~\mu$s. The parameters are setting as $\delta=\Omega_{p}=2\pi\times1~$MHz, $\Omega_{j}=0.05\Omega_{p}$, and $\Delta=2\pi\times300~$MHz.}
\end{figure}
\begin{equation}
H_{\textmd{sc}}=E_{1}|\psi_{1}\rangle\langle\psi_{1}|+E_{2}|\psi_{2}\rangle\langle\psi_{2}|
+E_{3}|\psi_{3}\rangle\langle\psi_{3}|,
\end{equation}
where the eigenvectors are given by
\begin{eqnarray}
|\psi_{1}\rangle&=&C_{11}|erg\rangle+C_{12}|rrg\rangle+C_{13}|reg\rangle,\nonumber\\
|\psi_{2}\rangle&=&C_{21}|erg\rangle+C_{22}|rrg\rangle+C_{23}|reg\rangle,\nonumber\\
|\psi_{3}\rangle&=&C_{31}|erg\rangle+C_{32}|rrg\rangle+C_{33}|reg\rangle\nonumber.
\end{eqnarray}
Through the transformation of representation, the basis vectors can be changed to states $|egg\rangle$, $|geg\rangle$, and $|\psi_{i}\rangle$ ($i=1,2,3$), and Eq.~(\ref{Es1}) changes to
\begin{equation}\label{Es2}
 H_{\textmd{I1}}=\left[
              \begin{array}{c|ccc|c}
              \rho_{1,1} & \tilde{\rho}_{1,2} & \tilde{\rho}_{1,3} & \tilde{\rho}_{1,4} & \rho_{1,5}\\
              \hline
              \tilde{\rho}_{1,2}^{*} & E_{1} & 0 & 0 & \tilde{\rho}_{2,5}\\
              \tilde{\rho}_{1,3}^{*} & 0 & E_{2} & 0 & \tilde{\rho}_{3,5}\\
              \tilde{\rho}_{1,4}^{*} & 0 & 0 & E_{3} & \tilde{\rho}_{4,5}\\
              \hline
              \rho_{1,5}^{*} & \tilde{\rho}_{2,5}^{*} & \tilde{\rho}_{3,5}^{*} & \tilde{\rho}_{4,5}^{*} & \rho_{5,5}
              \end{array}
\right],
\end{equation}
where
\begin{eqnarray}
\tilde{\rho}_{1,2}&=&\rho_{1,2}C_{11}+\rho_{1,3}C_{12}+\rho_{1,4}C_{13},\nonumber\\
\tilde{\rho}_{1,3}&=&\rho_{1,2}C_{21}+\rho_{1,3}C_{22}+\rho_{1,4}C_{23},\nonumber\\
\tilde{\rho}_{1,4}&=&\rho_{1,2}C_{31}+\rho_{1,3}C_{32}+\rho_{1,4}C_{33},\nonumber\\
\tilde{\rho}_{2,5}&=&\rho_{2,5}C_{11}^{*}+\rho_{3,5}C_{12}^{*}+\rho_{4,5}C_{13}^{*},\nonumber\\
\tilde{\rho}_{3,5}&=&\rho_{2,5}C_{21}^{*}+\rho_{3,5}C_{22}^{*}+\rho_{4,5}C_{23}^{*},\nonumber\\ \tilde{\rho}_{4,5}&=&\rho_{2,5}C_{31}^{*}+\rho_{3,5}C_{32}^{*}+\rho_{4,5}C_{33}^{*}.\nonumber
\end{eqnarray}

\begin{figure}
\centering\scalebox{0.32}{\includegraphics{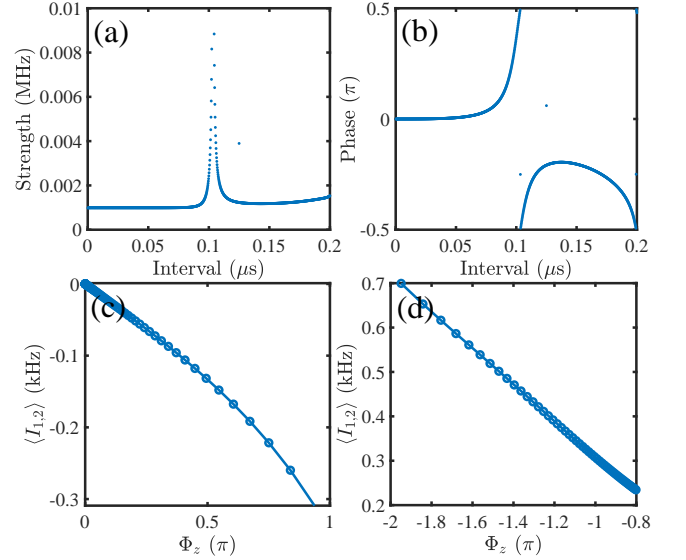}}
\caption{\label{Liu_four}The chiral motion of atomic excitation under a tight-binding model with $N=4$ arranged in a square. (a) and (b) respectively shows the average value of the effective coupling strength and induced phases under different time intervals. (c) and (d) The chiral current $\langle I_{1,2}\rangle$ of ground-state governed by the periodical Hamiltonian from the initial state $|eggg\rangle$. The corresponding time intervals are $\tau\in[0.001, 0.096]~\mu$s and $\tau\in[0.13, 0.199]~\mu$s, respectively. The other parameters are taken as $\delta=\Omega_{p}=2\pi\times1~$MHz, $\Omega=0.05\Omega_{p}$, and $\Delta=300\Omega_{p}$.}
\end{figure}

According to Eq.~(\ref{Es2}), the Raman transition between ground states $|egg\rangle$ and $|geg\rangle$ is assisted by multiple channels, the Hamiltonian of each channel can be written as
\begin{equation}
H_{\textmd{I1}}^{(1)}=\tilde{\rho}_{1,2}e^{-iE_{1}t}|egg\rangle\langle \psi_{1}|+\tilde{\rho}_{2,5}e^{iE_{1}t}|\psi_{1}\rangle\langle geg|+{\rm H.c.},
\end{equation}
\begin{equation}
H_{\textmd{I1}}^{(2)}=\tilde{\rho}_{1,3}e^{-iE_{2}t}|egg\rangle\langle\psi_{2}|+\tilde{\rho}_{3,5}e^{iE_{2}t}|\psi_{2}\rangle\langle geg|+{\rm H.c.},
\end{equation}
\begin{equation}
H_{\textmd{I1}}^{(3)}=\tilde{\rho}_{1,4}e^{-iE_{3}t}|egg\rangle\langle\psi_{3}|+\tilde{\rho}_{4,5}e^{iE_{3}t}|\psi_{3}\rangle\langle geg|+{\rm H.c.}.
\end{equation}

In the limit of large detunings with $E_{1}\gg \tilde{\rho}_{1,2(2,5)}$, $E_{2}\gg \tilde{\rho}_{1,3(3,5)}$, and $E_{3}\gg \tilde{\rho}_{1,4(4,5)}$, the excited states $|\psi_{i}\rangle$ can be adiabatically eliminated, and the effective coupling constant of ground-state transition $|geg\rangle\rightarrow|egg\rangle$ is
\begin{equation}
J_{12}=-J_{\textmd{eff}}^{12}e^{i\Phi_{12}}=\rho_{1,5}-\frac{\tilde{\rho}_{1,2}\tilde{\rho}_{2,5}}{E_{1}}
-\frac{\tilde{\rho}_{1,3}\tilde{\rho}_{3,5}}{E_{2}}-\frac{\tilde{\rho}_{1,4}\tilde{\rho}_{4,5}}{E_{3}}.
\end{equation}
Thus, the coupling strength and the induced phase read as
\begin{equation}
J_{\textmd{eff}}^{12}=|J_{12}|,\ \ \Phi_{12}=\frac{1}{\pi}\arctan[\frac{\textmd{Im}(J_{12})}{\textmd{Re}(J_{12})}].
\end{equation}
The same operations can be performed for the other two processes $|egg\rangle\rightarrow|gge\rangle$ and $|gge\rangle\rightarrow|geg\rangle$ and the effective Hamiltonian of the whole system can be obtained
\begin{eqnarray}
H_{\textmd{eff}}&=&-J_{\textmd{eff}}^{12}e^{i\Phi_{12}}|egg\rangle\langle geg|-J_{\textmd{eff}}^{23}e^{i\Phi_{23}}|geg\rangle\langle gge|\nonumber\\&&-J_{\textmd{eff}}^{31}e^{i\Phi_{31}}|gge\rangle\langle egg|+{\rm H.c.},
\end{eqnarray}
which can successfully lead to a chiral motion of atomic excitation with $\Phi_{12}+\Phi_{23}+\Phi_{31}=\pm\pi/2$.  Fig.~\ref{Phase} shows the numerical results of the effective couplings $J_{\textmd{eff}}^{ij}$ and induced phases $\Phi_{ij}$ ($ij=12,23,31$) between arbitrary ground states with different time intervals $\tau$, in which we calculate the original Hamiltonian shown as a $27\times27$ matrix for higher precision. To promise the results consistent with the ground-state dynamics, we keep the convergent results and discard the divergent results.
Fig.~\ref{Tau_para}(a)-(c) characterize the change of $\tilde{\tau}$ with $\Omega$, $\delta$, and $\Omega_{p}$, respectively. It illustrates that $\tilde{\tau}$ is mainly related to $\delta$ and $\Omega_{p}$. Meanwhile, on average, the homologous effective coupling strengths $J_{\textmd{eff}}$ are shown in Fig.~\ref{Tau_para}(d)-(f).
The corresponding $\tilde{\tau}$ here are ideal values which may be difficult to accurately control in experiment. To test the maneuverability, we reduce the accuracy of the time intervals $\tilde{\tau}_{ac(c)}$ shown in Fig.~\ref{current}(e) and \ref{current}(f) to three decimal places and plot the corresponding evolution as shown in Fig.~\ref{tau}. The chiral motion of atomic excitation can still be clearly observed.

As discussed before, the chiral motion can be achieved for triangle structure and a special current with definite direction can be reached by adjusting time intervals. Thus, a natural question to ask is what will happen for larger lattices under our protocol. For a square geometry, the effective coupling strength between the nearest neighbour atoms are basically the same with the periodical Hamiltonian read as
\begin{equation}
 H(t)=\left\{
              \begin{array}{lr}
              H_{1},\ \ t\in[0, T/4)\\
              H_{2},\ \ t\in[T/4, T/2)\\
              H_{3},\ \ t\in[T/2, 3T/4)\\
              H_{4},\ \ t\in[3T/4, T)\\
              \end{array}
\right.
\end{equation}
where
\begin{eqnarray}
H_{i}&=&\Omega_{i}|r_{i}\rangle\langle g_i|+\sum_{j=1}^{4}\Omega_{p}|r_{j}\rangle\langle e_j|+{\rm H.c.}
+\delta|g_{j}\rangle\langle g_j|\nonumber\\&&+\Delta|e_{j}\rangle\langle e_j|+\sum_{j<k}{\cal U}_{j k}|r_jr_k\rangle\langle r_jr_k|.
\end{eqnarray}
Under the same operations, the average value of the effective coupling strength $J_{\textmd{eff}}$ [$J_{\textmd{eff}}=1/4(J_{\textmd{eff}}^{12}+J_{\textmd{eff}}^{23}+J_{\textmd{eff}}^{34}+J_{\textmd{eff}}^{41})$]  and $\Phi$ [$\Phi=1/4(\Phi_{12}+\Phi_{23}+\Phi_{34}+\Phi_{41})$] according to different time intervals are shown in Fig.~\ref{Liu_four}(a) and \ref{Liu_four}(b).
By tuning the time interval $\tau$, the chiral motion of atomic excitation can be obtained for $\Phi_{z}\neq0,\pm2\pi$. As shown in Fig.~\ref{Liu_four}(c) and \ref{Liu_four}(d), the ground state current for bond $1\rightarrow2$ has been measured with $\tau\in[0.001, 0.096]~\mu$s and $\tau\in[0.13, 0.199]~\mu$s, respectively. The other parameters are taken as $\delta=\Omega_{p}=2\pi\times1~$MHz, $\Omega=0.05\Omega_{p}$, and $\Delta=300\Omega_{p}$. It is easy to find that the direction of the ground-state current is related to the sign of $\Phi_{z}$.
However, the chiral motion with each atom reaching the maximum population close to unity in clockwise or anticlockwise order only exists in the triangle structure with $\Phi_{z}=\pm\pi/2$.
\bibliography{pra-reply-2}
\end{document}